\documentclass[sn-nature]{sn-jnl}

\usepackage{graphicx}%
\usepackage{multirow}%
\usepackage{amsmath,amssymb,amsfonts}%
\usepackage{amsthm}%
\usepackage{mathrsfs}%
\usepackage[title]{appendix}%
\usepackage{xcolor}%
\usepackage{textcomp}%
\usepackage{manyfoot}%
\usepackage{booktabs}%
\usepackage{algorithm}%
\usepackage{algorithmicx}%
\usepackage{algpseudocode}%
\usepackage{listings}%
\usepackage{fullpage}

\theoremstyle{thmstyleone}%
\newtheorem{theorem}{Theorem}
\theoremstyle{thmstyletwo}%

\theoremstyle{thmstylethree}%

\newcommand{\todo}[1]{\textcolor{red}{{#1}}}
\newcommand{\reader}[1]{\textcolor{magenta}{{#1}}}

\def\method{%
  \textsc{Luna}%
  \futurelet\next\checknext
}

\def\checknext{%
  \ifx\next,%
  \else\ifx\next.%
  \else\ifx\next'%
  \else\ifx\next)%
  \else
    \ 
  \fi\fi\fi\fi
}
\usepackage[utf8]{inputenc} 
\usepackage[T1]{fontenc}    
\usepackage{hyperref}       
\usepackage{url}            
\usepackage{booktabs}       
\usepackage{amsfonts}       
\usepackage{nicefrac}       
\usepackage{microtype}      
\usepackage{xcolor}         
\usepackage{tikz}
\usepackage{pgfplots}
\usepackage{subcaption}
\usepackage{wrapfig}
\usepackage{enumitem}
\usepackage{multirow}
 \usepackage{multirow}
  \usepackage{booktabs,siunitx}
\usepackage{pifont}
 \usepackage{diagbox}
\usepackage{bm}
\usepackage{bbm}
\usepackage{amssymb}
\usepackage{amsmath}
\usepackage{mathtools}
\usepackage[nameinlink]{cleveref}
\usepackage{dutchcal} 
\usepackage{amsmath, bm}
\newcommand{\matr}[1]{\mathbf{#1}}
\usepackage{bold-extra}

\hypersetup{
	colorlinks=true,
	linkcolor=myblue,
	citecolor=myblue,
	urlcolor=mygray,
}

\definecolor{myblue}{RGB}{0,0, 150}
\definecolor{mygray}{RGB}{90,90,110}
\begin{document}

\title[Article Title]{\centering Ultrasound Lung Aeration Map via \\ Physics-Aware Neural Operators}

\author[1]{\fnm{Jiayun} \sur{Wang}}\email{peterw@caltech.edu}
\equalcont{Equal contribution.}

\author[2,3]{\fnm{Oleksii} \sur{Ostras}}\email{oleksii@email.unc.edu}
\equalcont{Equal contribution.}

\author[2,3]{\fnm{Masashi} \sur{Sode}}\email{msode@email.unc.edu}

\author[1]{\fnm{Bahareh} \sur{Tolooshams}}\email{btoloosh@caltech.edu}

\author[1]{\fnm{Zongyi} \sur{Li}}\email{zongyili@caltech.edu}

\author[4]{\fnm{Kamyar} \sur{Azizzadenesheli}}\email{kamyara@nvidia.com}

\author[2,3]{\fnm{Gianmarco F.} \sur{Pinton}}\email{gia@email.unc.edu}
\equaladv{Equal advising.}

\author[1]{\fnm{Anima} \sur{Anandkumar}}\email{anima@caltech.edu}
\equaladv{Equal advising.}

\affil[1]{\orgdiv{Department of Computing and Mathematical Sciences}, \orgname{California Institute of Technology}, \orgaddress{\street{1200 E California Blvd}, \city{Pasadena}, \postcode{91125}, \state{CA}, \country{United States}}}

\affil[2]{\orgdiv{Department of Biomedical Engineering}, \orgname{University of North Carolina at Chapel Hill}, \orgaddress{\street{103 South Building}, \city{Chapel Hill}, \postcode{27514}, \state{NC}, \country{United States}}}

\affil[3]{\orgdiv{Department of Biomedical Engineering}, \orgname{North Carolina State University}, \orgaddress{\street{Campus Box 7625}, \city{Raleigh}, \postcode{27695}, \state{NC}, \country{United States}}}

\affil[4]{ \orgname{NVIDIA}, \orgaddress{\street{2788 San Tomas Express Way}, \city{Santa Clara}, \postcode{95051}, \state{CA}, \country{United States}}}


\abstract{Lung ultrasound is a growing modality in clinics for diagnosing and monitoring acute and chronic lung diseases due to its low cost and accessibility. Lung ultrasound works by emitting diagnostic pulses, receiving pressure waves and converting them into radio frequency (RF) data, which are then processed into B-mode images with beamformers for radiologists to interpret. However, unlike conventional ultrasound for soft tissue anatomical imaging, lung ultrasound interpretation is complicated by complex reverberations from the pleural interface caused by the inability of ultrasound to penetrate air. The indirect B-mode images 
make interpretation highly dependent on reader expertise, requiring years of training, which limits its widespread use despite its potential for high accuracy in skilled hands.

To address these challenges and democratize ultrasound lung imaging as a reliable diagnostic tool, we propose \method (the Lung Ultrasound Neural operator for Aeration), an AI model that directly reconstructs lung aeration maps from RF data, bypassing the need for traditional beamformers and indirect interpretation of B-mode images. \method uses a Fourier neural operator, which processes RF data efficiently in Fourier space, enabling accurate reconstruction of lung aeration maps. 
From reconstructed aeration maps, we calculate lung percent aeration, a key clinical metric, offering a quantitative, reader-independent alternative to traditional semi-quantitative lung ultrasound scoring methods.
The development of \method involves synthetic and real data: We simulate synthetic data with an experimentally validated approach and scan ex vivo swine lungs as real data. 
Trained on abundant simulated data and fine-tuned with a small amount of real-world data, \method achieves robust performance, demonstrated by an aeration estimation error of 9\% in ex-vivo swine lung scans. 
We demonstrate the potential of directly reconstructing lung aeration maps from RF data, providing a foundation for improving lung ultrasound interpretability, reproducibility and diagnostic utility.
}




\maketitle
 \section{Introduction}

Lung ultrasound (LUS) is an important non-invasive real-time imaging modality widely used for diagnostics and monitoring of lung disease in its acute and chronic phases, such as respiratory and extrapulmonary diseases \cite{bouhemad2015ultrasound,lacedonia2021role,manolescu2020ultrasound}. Compared to X-ray imaging and computed tomography (CT), lung ultrasound has inherent advantages: it is non-ionizing, portable, low-cost and suitable for frequent or bedside monitoring. Furthermore, the unique acoustic interaction between soft tissue and air provides distinct contrast mechanisms, potentially offering complementary diagnostic information to X-ray and CT imaging -- expert users can achieve a sensitivity and specificity of 90\% to 100\% for disorders including pleural effusions, lung consolidation, pneumothorax and interstitial syndrome \cite{lichtenstein2014lung}. 

Despite its promise, barriers are limiting the widespread adoption of lung ultrasound because interpreting lung ultrasound images usually requires years of experience \cite{see2016lung,vitale2016comparison}.
Unlike other diagnostic ultrasound (e.g. fetal and abdominal ultrasound \cite{whitworth2015ultrasound,mattoon2014abdominal}), lung ultrasound images primarily rely on the interpretation of artifacts created by reverberations at the tissue-air interface \cite{demi2022}, rather than direct visualization of lung structures \cite{soldati2020}. This indirect imaging mechanism makes clinical interpretation highly dependent on the operator’s expertise and the imaging system’s parameters (e.g., central frequency, focal depth gain settings), leading to significant variability and insufficient interobserver agreement \cite{pivetta2018sources,gomond2020effect,gravel2020interrater}. Furthermore, current delay-and-sum beamforming methods, designed for soft tissue imaging, are not optimized for the unique acoustic physics of lung tissue, limiting diagnostic accuracy \cite{ostras2023development}.

\begin{figure}[t!]
\vspace{4em}
    \centering
    \includegraphics[width=\columnwidth]{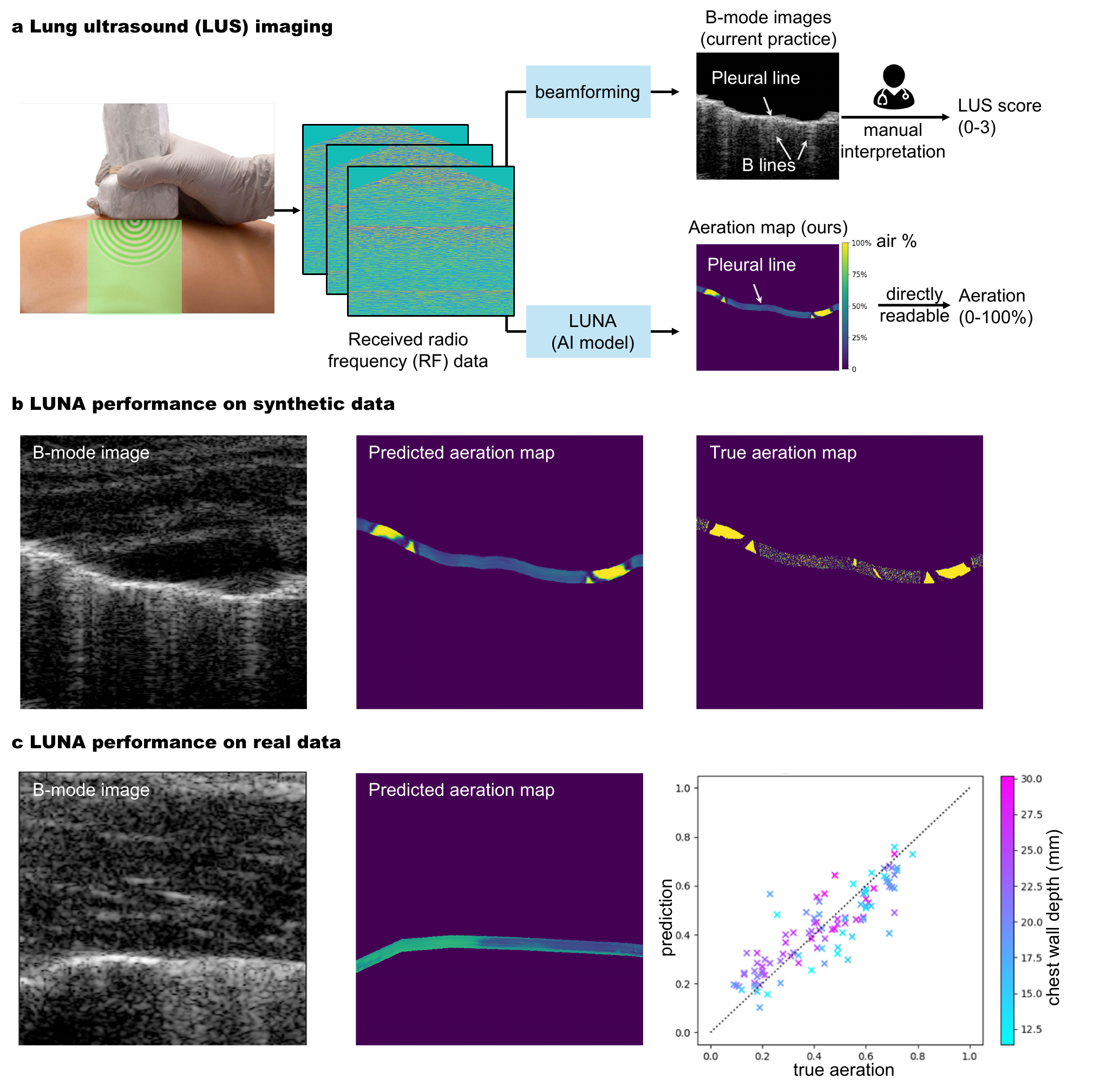} 
    \caption{\small \textbf{Overview: \textsc{Luna} (the Lung Ultrasound Neural operator for Aeration) reconstructs lung aeration maps from ultrasound radio-frequency (RF) data}.  {\bf a}, The lung ultrasound (LUS) imaging process: Ultrasound devices scan the lung and 
 the RF data is fed to \textsc{Luna} for reconstructing a lung air-tissue map, or aeration map, which is human-interpretable. The aeration map is used to estimate lung percent aeration, a critical clinical outcome for evaluation, monitoring and diagnostics \cite{lee2020lung,kalkanis2022lung}. 
{\bf Existing v.s. our approach:} The current practice, B-mode lung image (upper right) requires manual interpretation of artifacts like B lines to assign LUS score \cite{volpicelli2012international}, which has high variations due to the difficulty of identifying artifacts created by the complex wave propagation physics in the lung. LUS score is also coarse, image-level and semi-quantitative.
{\bf Our approach} (lower right): We reconstruct lung aeration maps that directly depict tissue-air maps, from which clinicians can directly read pixel-level aeration. The method is reproducible and provides quantitative two-dimensional aeration distribution.
{\bf b,} \textsc{Luna}  performance on synthetic lung ultrasound data. Pixel-level aeration can be read from the predicted aeration map, which is visually similar to the true aeration map.
{\bf c,} \textsc{Luna}  performance on real {\it ex vivo} lung ultrasound data. The average aeration prediction error is $9.4\%$, which well outperforms the sensitivity of the current scoring system. }
    \label{fig:teaser}
\end{figure}

Recent advances in machine learning have introduced automated methods that assist clinicians for lung ultrasound image interpretation and diagnostic purposes \cite{xing2023,sagreiya2023}, such as identifying horizontal (A-lines) and vertical (B-lines) artifacts or directly classifying diseases like COVID-19 and pneumonia from brightness-mode (B-mode) images \cite{horry2020covid,born2020pocovid,roy2020deep}. However, these approaches that directly learn from B-mode images instead of raw radiofrequency (RF) data face several challenges: 1) the information related to lung morphology at mesoscopic/alveolar level is lost in the beamforming process \cite{mento2024}; 2) such models do not generalize well to different ultrasound devices as each device has a setting with different imaging parameters (dynamic range, time gain compensation) and quality of the resulting B-mode images also vary \cite{xu2023}. We provide examples in \cref{fig:tgc} where changes of time gain compensation result in changes in artifacts of B-mode images for final outcome prediction.

{\bf Our approach:} We take a fundamentally different approach to lung ultrasound analysis. Rather than interpreting on B-mode images, either manually or automatically \cite{lichtenstein2014lung,lee2014lung,baloescu2020automated}, we propose 
an AI model, \method (the Lung Ultrasound Neural operator for Aeration), that directly reconstructs lung aeration maps from delayed ultrasound radiofrequency (RF) data, bypassing the traditional beamforming process. Delayed RF data, the input data to \method,  is a space-time representation that accounts for the time-of-flight delays of raw wave propagation data in the lung. \method offers two significant advantages: 
1) By directly processing raw RF data, it preserves diagnostic information otherwise lost in beamforming and reduces variability caused by device and parameter differences. 
2) By reconstructing lung aeration maps, it eliminates the need for interpreting B-mode artifacts, providing a more direct and quantitative representation of lung disease.
Notably, the reconstructed lung aeration map enables the calculation of the lung percent aeration, a critical clinical measurement for diagnosis \cite{lee2020lung,kalkanis2022lung}. 
3) The reconstructed lung aeration maps provide a quantitative analysis of lung status, while the existing clinically-adopted lung ultrasound score system, LUS score  \cite{volpicelli2012international}, only provides a coarse and semi-quantitative index for assessing lung aeration. \method could establish a quantitative link between ultrasound propagation and the disease state of the lung. 

\method is designed to address two key challenges in reconstructing lung aeration maps from RF data:
1) {\bf The complexity of ultrasound propagation in the lung}. The highly reflective tissue-air interface and multiple scattering make the inverse problem ill-posed, necessitating a learning-based approach capable of efficiently capturing subtle changes in RF data. Our \method, based on Fourier neural operators \cite{li2021fourier}, learns maps between function spaces and features parameterized directly in Fourier
space, capturing subtle RF data variations and enabling efficient extraction of diagnostically relevant features across spatial and temporal scales.
2) {\bf The difficulty of collecting real paired data}. Collecting real paired RF-lung aeration map data is time-consuming. We use an experimentally validated simulation approach, Fullwave-2 \cite{pinton2021fullwave}, which solves the full-wave equation to model nonlinear wave propagation, frequency-dependent attenuation and density variations. Its ability to accurately simulate ultrasound propagation in the lung's complex acoustic environment is crucial for understanding the relationship between RF data and lung aeration maps. 
We also acquire real lung ultrasound data by scanning fresh swine lung tissues, whose ground-truth aeration is measured for validating purposes.
\method is thus trained on the abundant simulated data and fine-tuned on a small number of real data samples. 
\method achieves 9.4\% error on predicting percent aeration for real ex-vivo swine lung, which outperforms the current lung ultrasound scoring system \cite{volpicelli2012international} and has been considered satisfactory by experienced radiologists involved in the study. 

To summarize, the proposed \method has two main contributions:
 {\bf 1) Pioneering ultrasound lung aeration map reconstruction.} This work is the first to reconstruct lung aeration maps from ultrasound RF data, enabling accurate percent aeration estimation and supporting future diagnostic applications.
{\bf 2) Effectiveness of training on simulation data and its generalizability to real scenarios.} 
We demonstrate the power of learning wave propagation physics in the lung via physics simulation: Training on abundant simulation data, our model demonstrates satisfactory generalizability on \textit{ex vivo} data.

\begin{figure}[t!]
    \centering
   \includegraphics[width=\columnwidth]{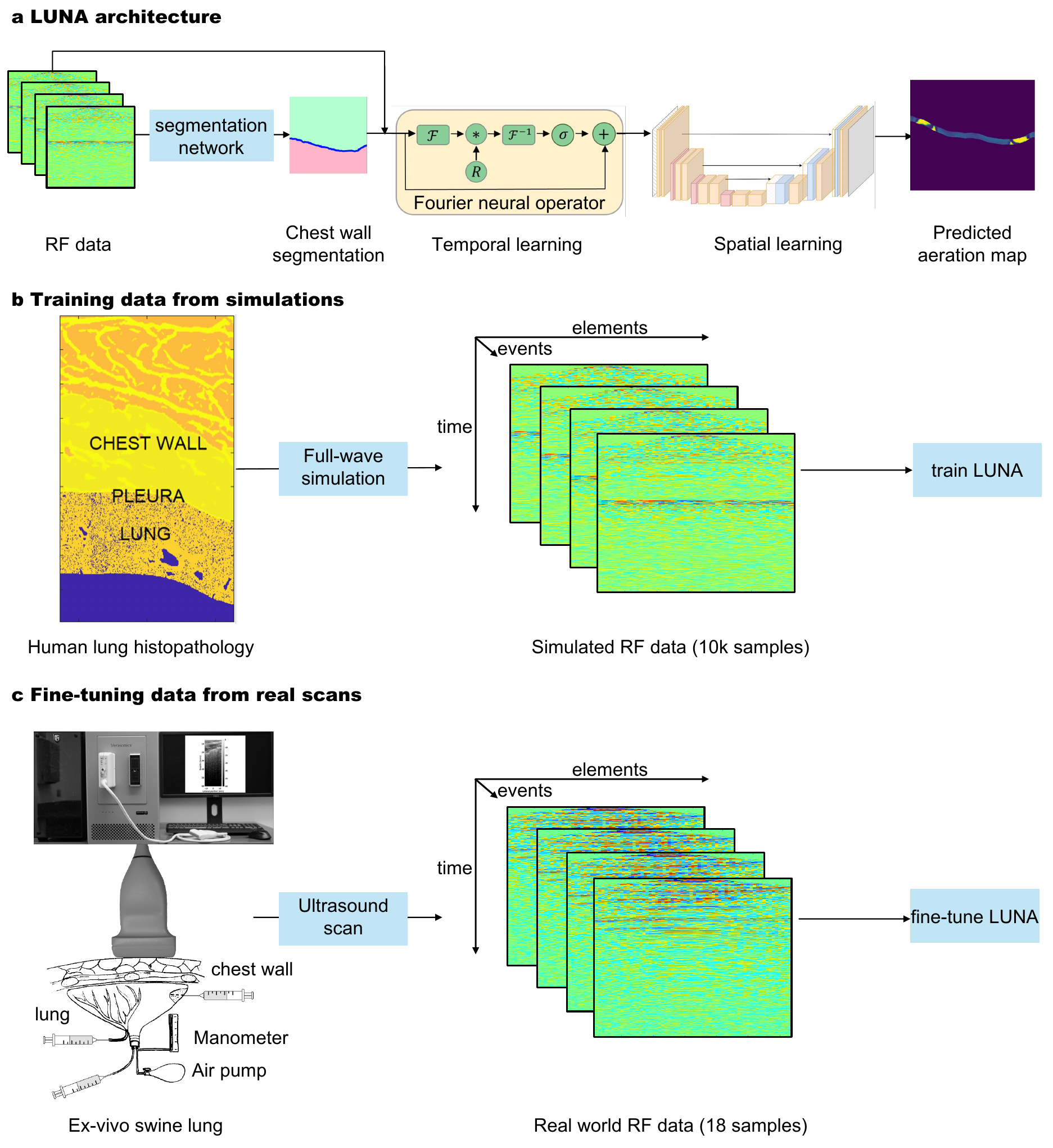} 
    \caption{ \small
    {\bf \method architecture and training/fine-tuning data.} 
    {\bf a,} Pipeline: During training, ultrasound RF data undergoes data augmentation and is processed by the physics-aware neural operator, \method. \method comprises two main components: a temporal Fourier neural operator, which processes the temporal dimension of RF data (corresponding to the output's depth dimension) and a spatial network, which captures lateral interactions in the RF data (corresponding to the output's lateral dimension). The model outputs a reconstructed aeration map, where each pixel value represents the aeration percentage.  
    {\bf b}, Generation of synthetic data, which is used for training \method. Left: Combined aeration map comprised of the tissue-specifically segmented body wall (top) and underlying lung deformed to conform to its internal surface which models a realistic pleural interface. 
   Middle: Stack of numerically simulated raw RF data of 128 transmit-receive events visualized as the amplitude of received backscattered signal in form receiver-time. Fullwave-2 \cite{pinton2021fullwave}, a fullwave model of the nonlinear wave equation, is used as the simulation tool. 
    10k samples are used to train \method.
       {\bf c,} Acquisition of real data, which is used for training \method. 
       Scanning of fresh porcine lungs of known aeration (displacement method) through chest wall fragment in the water tank (\textit{ex vivo}) using a programmable ultrasound machine and linear transducer. 18 samples are used to fine-tune \method.
    }
    \label{fig:data}
\end{figure}

\section{Results}


\bmhead{\method, a deep neural operator framework for lung aeration map reconstruction}
We implement \method (the Lung Ultrasound Neural operator for Aeration), a deep learning framework for reconstructing lung aeration map $\rho^A$ from the radio frequency (RF) data/acoustic pressure $p$ received by the ultrasound transducer (\cref{fig:teaser}a). 
 Lung ultrasound imaging involves placing the ultrasound probe at the body surface and using the RF data received at the surface for diagnostics and monitoring of lung disease. 
The inverse problem of reconstructing lung aeration map  ${\rho}^A$  (obtained from reconstructed lung density image) from RF signal $p$ is challenging due to the complex, high nonlinear wave propagation physics and the ill-pose nature of the problem.
\method learns to approximate the inverse operator with a non-linear parameterized model $f_{\theta}$
\begin{align}
    \hat{\rho}^A = f_{\theta}(p)
\end{align}

\underline{Comparison to Existing Beamforming Method}. 
In the current practice, the high-dimensional ultrasound RF data is beamformed and compressed into a 2D B-mode image (\cref{fig:teaser}a-upper right) \cite{gargani2014lung,lichtenstein2014lung}, which is an indirect way to infer the complicated wave propagation physics in the lung and usually requires years of experience for radiologists to effectively interpret such images \cite{see2016lung,vitale2016comparison}.
In contrast, \method's input is the delayed ultrasound RF data, a space-time representation that accounts for the time-of-flight delays of wave propagation in the lung. \method bypasses the beamforming which compresses the RF data and directly estimates the lung aeration map, facilitating easy and interpretation-free clinical uses (\cref{fig:teaser}a-lower right).

\underline{Neural Operator Framework is Suitable for the Task}. \method's architecture (\cref{fig:data}a) is based on the Fourier neural operator (FNO) \cite{li2021fourier}, a deep neural architecture designed to learn maps between function spaces by parameterizing the integral kernel directly in Fourier space. 
Learning to extract diagnostically relevant features in the image space from the underlying wave propagation physics requires methods that span different time and spatial scales. Ultrasound data measured at the transducer surface is a function of space $\times$ time, but images must represent the correct information in a space $\times$ space by extracting relevant information at different moments in time and transferring them to the correct location in space. The location of this information is not known \textit{a priori}. Resolution agnostic approaches, such as FNO\cite{li2021fourier}, are ideally suited to this task since information is intrinsically learned in Fourier space which does impose specific spatial or temporal constraints on the location of information representation. 

Taking ultrasound RF data $p$ as the input, \method first predicts the chest wall v.s. lung tissue segmentation. The segmentation is then combined with the RF data to reconstruct the lung aeration map $\hat{\rho}^A$. Implementation details including the deep learning architecture, training strategies, hyperparameters and post-processing are provided in  Section \ref{sec:implemntation}.
The reconstructed lung aeration map is human-interpretable and can enable many downstream clinical outcomes. Specifically, we consider an important clinical sign, the percent lung aeration $\gamma$, which can be calculated from the reconstructed lung aeration map $\hat{\rho}^A$.


\underline{Two-Stage Learning of \method}. 
As discussed in the introduction, we adopt a two-stage training and \method is firstly trained on the abundant simulated data and fine-tuned on a small number of real data samples, due to the cost of obtaining real data. 
Specifically, The first stage of the two-stage learning is to train \method on $10,150$ simulated lung ultrasound RF data/aeration map pairs. In silico lung aeration map reconstruction is validated with the ground truth to ensure satisfactory performance. In the second stage, \method is fine-tuned on 18 {ex-vivo} data samples. The fine-tuned \method is then validated on 103 {ex-vivo} data samples. Both stages involve data augmentation for improved robustness of \method. We present more details on the data below.

\bmhead{\method is trained on experimentally-validated simulation data of lung ultrasound propagation} 
Training a machine learning surrogate model for the inverse problem of lung ultrasound wave propagation requires paired data of lung aeration maps and their corresponding RF signals, enabling the model to learn the underlying mapping. 
Obtaining sufficient real paired data {\it ex-vivo} or {\it in-vivo} can be time-consuming, 
we thus train \method on the experimentally validated simulated lung ultrasound data and fine-tune the model on real {\it ex-vivo} data.
The simulation tool, Fullwave-2 \cite{pinton2021fullwave}, solves the full-wave equation~\cite{pinton2009}. In addition to modeling the nonlinear propagation of waves, it describes arbitrary frequency-dependent attenuation and variations in density. The unique ability to accurately model ultrasound propagation in the complex acoustic environment provided by the lung and body wall is a key innovation required for reconstructing lung aeration maps from ultrasound RF data.
In the following, we discuss how we obtain the simulated data.


\underline{Simulated Data Generation}. 
Maps of acoustical properties were created by combining segmented axial anatomical images of the human chest wall (Visible Human Project, resolution 330 $\mu$m) with high-resolution histological images of healthy swine lung tissue (5 $\mu$m thickness, 0.55 $\mu$m resolution), similar to \cite{ostras2023} (\cref{fig:data}b).
First, anatomical chest wall maps and microscopic lung structures were merged to produce acoustical maps. Lung aeration, the ratio of air to non-air in the lung, was quantified from binary-segmented histological images, creating aeration maps where 1 represents air and 0 represents non-air ($\rho^A: \mathbb{Z}^2 \to {0, 1}$). The lung histopathology processing steps are illustrated in \cref{fig:sup_hist} in the Supplementary. 
Next, simulations of diagnostic ultrasound pulses were conducted using a clinically relevant setup (linear transducer, focused transmit sequence at 5.2 MHz). Corresponding RF signals ($p$) received by the transducer were collected. These simulations were performed with the Fullwave-2 acoustic simulation tool \cite{pinton2021fullwave}, which has been experimentally validated for reverberation, phase aberration and tissue-specific acoustic effects such as attenuation, scattering and absorption \cite{soulioti2021, pinton2011attenuation}. Details are available in Section \ref{sec:lussim}.



\bmhead{\method's performance on lung aeration map reconstruction, { in silico}}

We report numerical results and visualizations of the lung aeration map reconstruction and percent aeration estimation {\it in silico}. We also analyze \method's performance grouped by different factors, including ground-truth percent aeration and chest wall depth. 

\underline{Evaluation Metrics}. The reconstructed lung aeration map by \method is a 2D image, which can be compared with the ground-truth lung aeration map with metrics including PSNR (peak signal-to-noise ratio) and SSIM (structural similarity index measure) {\it in silico}.
As an important clinical outcome, lung aeration $\gamma$ can be obtained by averaging lung aeration map $\rho^A$:
\begin{align}
\label{eq:aeration}
    \gamma = \frac{ \sum_{i}^{H} \sum_{j}^{W} \rho^A(i, j) }{HW}
\end{align}
where $H, W$ denotes the spatial size of the aeration map. We use the absolute error between the predicted lung percent aeration $\hat{\gamma}$ and the ground-truth percent aeration ${\gamma}$: $|\hat{\gamma}-\gamma|$, to evaluate the performance of the percent aeration estimation performance.

\begin{figure}[t!]
    \centering
   \includegraphics[width=\columnwidth]{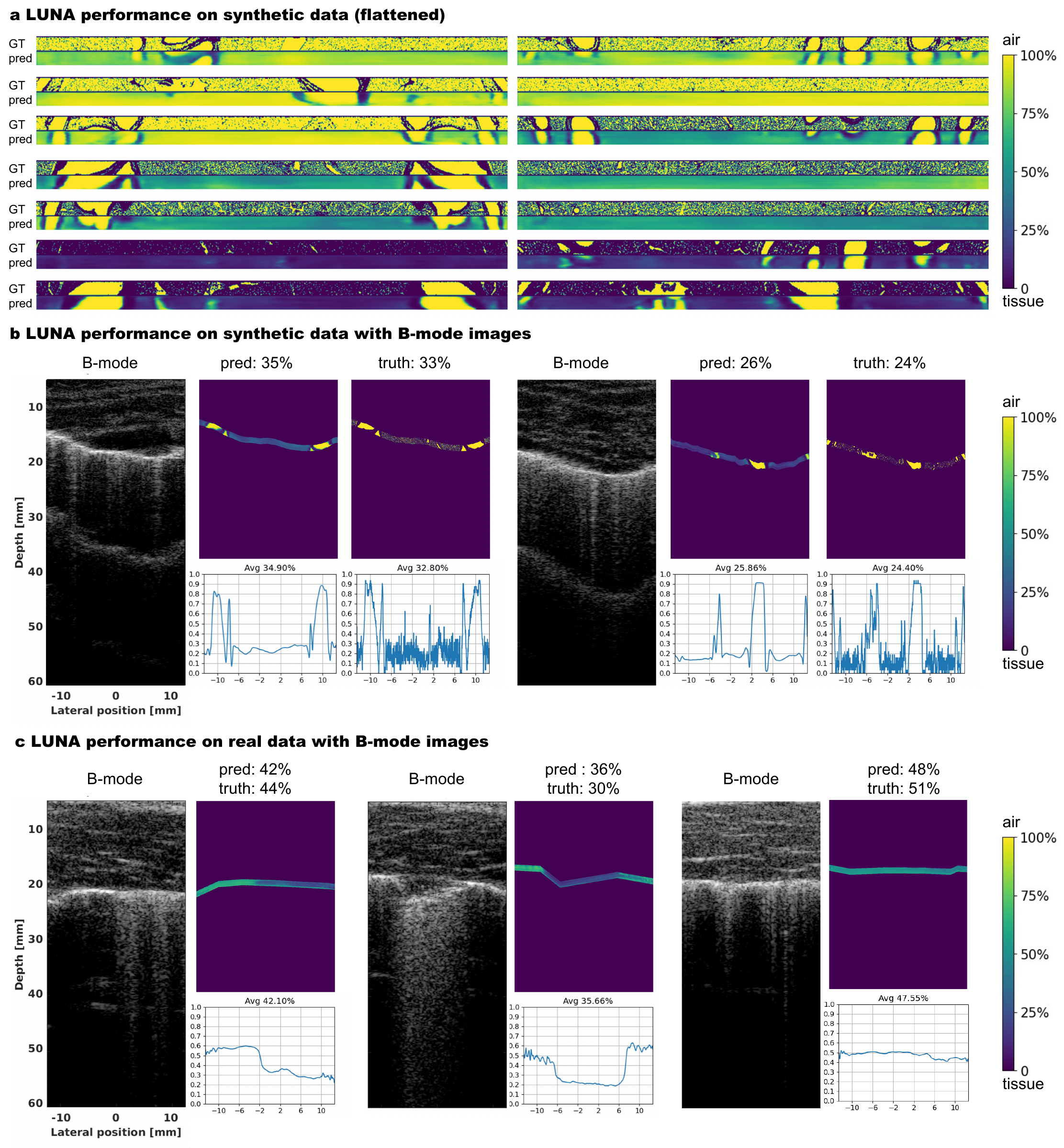} 
    \caption{ \small
    {\bf Visualization of lung ultrasound B-mode images and aeration maps, \textit{in silico} and \textit{ex vivo}.}
    {\bf a,}  Ground-truth and reconstructed aeration maps in \textit{in silico}, showing close visual similarity. Aeration maps are flattened for visualization purposes. 
    {\bf b/c,}  Reconstructed aeration maps overlaid on B-mode images for \textit{in silico} (b) and \textit{ex vivo} (c) experiments. The reconstructed aeration maps closely align with the B-mode images, effectively capturing aeration changes. We also provide \textit{in silico} ground-truth aeration maps to which the predictions are similar (b).
    The lower right plot in each subfigure depicts the reconstructed 1D percent aeration curve. 
    }
    \label{fig:ml_results}
\end{figure}

\begin{figure}[htb]
    \centering
   \includegraphics[width=\columnwidth]{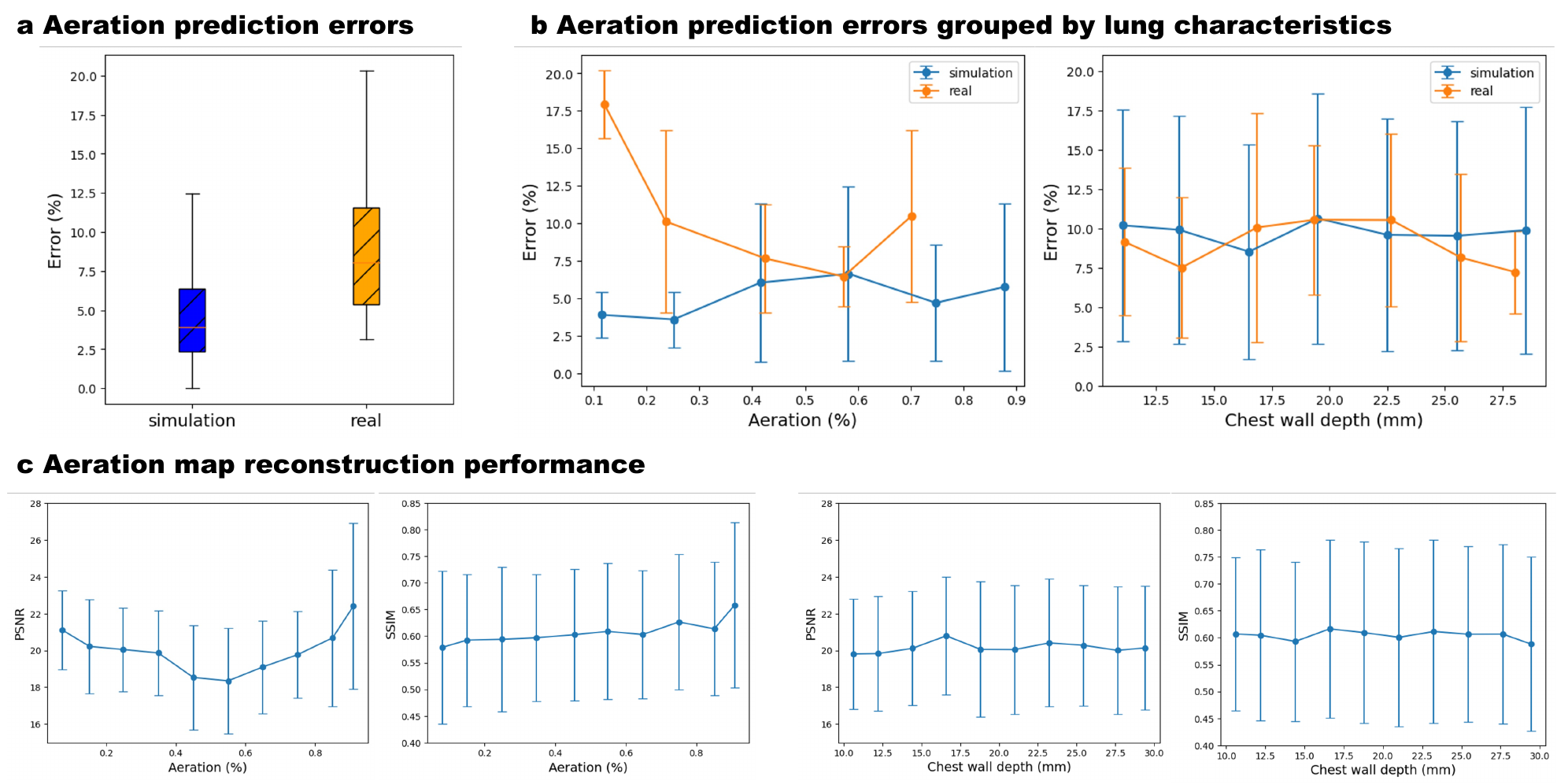}
    \caption{  \small
{\bf \method performance, \textit{in silico} and \textit{ex vivo}.}
    {\bf a,} \method percent aeration prediction error of \textit{in silico} and \textit{ex vivo} data. The error gap between them is 4.2\%. 
    {\bf b,} \method percent aeration prediction error grouped by ground-truth percent aeration and chest wall depth, \textit{in silico} and \textit{ex vivo}. Both sets have consistent errors for different aeration and chest wall depth, indicating the \method's robustness to such lung characteristics. 
    {\bf c,} \method aeration map 2D reconstruction performance (PSNR and SSIM) grouped by ground-truth percent aeration (left two subfigures) and chest wall depth (right two subfigures) \textit{in silico}. The performance is consistent for samples with different percent aeration and chest wall depth.
    }
    \label{fig:ml_results2}
\end{figure}

\underline{Lung Aeration Map Reconstruction Performance.} The average PSNR and SSIM on the evaluation set of in silico data are $20.10 \pm 3.51$ dB and $0.605 \pm 0.128$, respectively, demonstrating acceptable performance of \method in reconstructing 2D aeration maps. Visual comparisons of flattened reconstructed and ground-truth aeration maps (\cref{fig:ml_results}a) and comparisons of B-mode images with reconstructed aeration maps (\cref{fig:ml_results}b) qualitatively validate \method's accuracy. Localized alignment between B-mode image features (A-lines and B-lines) and reconstructed aeration maps further supports the reconstruction's precision.
We also report the average and standard deviation of PSNR and SSIM grouped by ground-truth percent aeration and chest wall depth, as in \cref{fig:ml_results2}e-f, respectively. The lung aeration map reconstruction performance is consistent over different percent aerations and chest wall depths, suggesting the performance of \method was insensitive towards lung properties including percent aeration and chest wall depth. 


\underline{Percent Aeration Estimation Performance.} 
As a validation of \method's performance in estimating the clinical outcome, percent aeration, 
\cref{fig:ml_results2}c depicts the box plot of the percent aeration estimation error in silico. The mean and standard deviation of the percent aeration estimation error is $5.2\%$ and $4.6\%$ and the highest percent aeration error is 12.5\%. 
\cref{fig:ml_results2}d depicts the percent aeration errors grouped by ground-truth percent aeration and chest wall depth, where the in silico percent aeration estimation error is consistent with varying aeration and chest wall depth.






\bmhead{\method has strong generalizability to real {ex-vivo} data} After training on simulated data, we fine-tune \method on a small number of real data (the two-stage training mentioned before) and show its generalizability on the real data. In the following, we discuss the real ex-vivo data acquisition and \method's performance on the real data. 

\underline{Real Ex-Vivo Data}.
Fresh tissues of 12 swine were hand-held scanned in a water tank using linear transducers and a research ultrasound system, with the same setting as the simulation data. 
Units obtained out of the right lung and caudal lobe of the left lung were employed in the calculation of its ground-truth percent aeration, while the left cranial lobe was used for scanning (\cref{fig:data}c). The ground-truth percent aeration can be compared with the predicted one from \method, although the ground-truth lung aeration map is unavailable due to the implausibility of obtaining it in reality. The depth of the reconstructed lung aeration map is set to be 2.6 ultrasound wavelength, considering the ultrasound penetration abilities. More details are provided in Section \ref{sec:realdata}. 

The data used in the study is summarized in Table \ref{tab:data_comp}. The simulated and real data have a relatively low domain gap in terms of percent aeration, chest wall depth distribution (\cref{fig:sup_data}a in the Supplementary), and B-mode image similarity (\cref{fig:sup_data}b in the Supplementary). The small gap between the two settings allows the robustness of \method's performance on both synthetic and real data.

\begin{table}[b]
\centering
\footnotesize
\caption{Lung ultrasound data used in the study. }
    \setlength{\tabcolsep}{2pt}
  \begin{tabular}{llllllll}
 \toprule

                                & \multicolumn{4}{c}{Simulation Set}                                            & \multicolumn{3}{c}{Ex-vivo Set}                 \\  \cmidrule{2-8} 
                                & Training                 & Validation               & Evaluation       &Sum        & Fine-Tuning            & Evaluation     &Sum         \\ \midrule
Number of data samples          & 10,150 & 1,450 & 2,900 &14,500 &  18 & 103 &121 \\ \midrule
Chest wall depth (mm), average$\pm$SD &      20.1$\pm$5.8          &   20.2$\pm$5.7                       &     20.3$\pm$5.7                     &          \slash              & 21.5$\pm$5.6 &  21.1$\pm$5.6& \slash                         \\ \midrule
Aeration (\%), average$\pm$SD         &       59.8$\pm$28.3                   &    53.8$\pm$24.1                      &60.1$\pm$25.0 & \slash &   33.4  & 43.2 &    \slash          \\
\bottomrule
\end{tabular}

          \label{tab:data_comp}
\end{table}

\underline{Results}. We report the evaluation results of \method on ex vivo swine lungs. \cref{fig:ml_results2}b depicts the predicted and the ground-truth percent aeration. \method achieves a lower than $10\%$ average percent aeration estimation error on the ex-vivo data. The data samples gather around the perfect prediction line, indicating the robust ex vivo performance of \method.

\underline{Percent Aeration Estimation Performance.} \cref{fig:ml_results2}c depicts the box plot of the percent aeration estimation error ex vivo. The mean and standard deviation of the percent aeration estimation error is $9.4\%$ and $5.4\%$, with the highest error over all cases being 20.3\%. 
The ex vivo percent aeration estimation performance is just slightly higher (4.2\%) than the in silico performance, indicating the good generalizability of \method in real data. 
\cref{fig:ml_results2}d depicts the ex vivo percent aeration estimation error is consistent with varying aeration and chest wall depth, where we see \method is insensitive to the lung properties like aeration and chest wall depth, ex vivo. Additionally, the error gap between in silico and ex vivo data is also small.


\underline{Visualization.} Although the ground-truth lung aeration maps {\it ex-vivo} are unavailable, we provide 
visual comparisons of B-mode images and reconstructed lung aeration maps in \cref{fig:ml_results}c—the localized alignment between B-mode image features (A-lines and B-lines) and reconstructed aeration map validates the accuracy of the aeration map reconstruction.

\underline{Runtime.} The processing time for each case is 0.11 seconds on an NVIDIA 4090 GPU, which means the framework can achieve an ideal aeration map display rate at 9 Hz.


\section{Discussion}
In this work, we introduce \method, a machine learning framework that, for the first time, reconstructs lung aeration maps directly from the lung ultrasound data.  
 Unlike traditional approaches that rely on human interpretation of B-mode images, \method bypasses this step and addresses challenges such as the expertise required to effectively interpret the complex physics of wave propagation in the lung and inconsistencies introduced by varying beamforming settings across ultrasound machines.
Additionally, \method improves the efficiency of lung disease screening, which can potentially improve clinical interpretability, reproducibility and diagnostic capabilities. 

\method also provides a robust quantitative evaluation of 2D lung aeration distribution, addressing limitations of current clinical practices. Currently, clinicians rely on semi-quantitative lung ultrasound scoring systems, LUS score, ranging from 0 to 3 \cite{volpicelli2012international}, which are prone to high individual variability, lack repeatability and offer only one coarse estimation score of lung conditions per scan. In contrast, \method’s reconstructed aeration maps provide pixel-level 2D lung aeration information derived directly from raw ultrasound signals per scan—data that is currently inaccessible to clinical interpretation. The advancement lays the groundwork for a more precise and reproducible method of linking ultrasound imaging to lung pathology, enabling improved diagnostic accuracy and clinical decision-making.

Our approach differs significantly from existing works that analyze lung ultrasound B-mode images to extract features like A-lines and B-lines \cite{zhao2024detection,hou2020interpretable} or classify diseases such as COVID-19 and pneumonia \cite{horry2020covid,born2020pocovid,roy2020deep}. Instead, \method operates on delayed RF data, preserving frequency domain information and minimizing variations introduced by imaging settings like dynamic range and time gain compensation. This distinction allows for more consistent and precise analyses across diverse clinical setups.

A key technical strength of \method is its ability to generalize from simulated data to real lung ultrasound data, a notable achievement given the challenges in lung imaging and the lack of paired real-world RF and aeration map data. Using a full-wave acoustic pressure field simulator, Fullwave-2 \cite{pinton2021fullwave}, we trained \method on simulated lung aeration maps and verified its robust performance in clinically relevant setups, including ex vivo swine lung experiments.

While Fullwave-2 \cite{pinton2021fullwave} effectively models ultrasound propagation in the human body, the feasibility of a learning-free iterative solver for the inverse problem remains questionable. Such solvers have not been explored, likely due to two main challenges:
1) {Ill-posedness and instability}. RF data is captured only at the transducer surface, providing partial wave information. This makes it difficult to impose the full-wave equation as a constraint, resulting in unstable optimization.
2) {Unrealistic computational demands}. Unlike geophysical applications \cite{virieux2009overview,li2020recent}, the lung’s spatial complexity, air-filled structures and phenomena such as reverberation, multiple scattering and nonlinearity render optimization-based solvers computationally infeasible.
\method addresses these challenges by using machine learning to directly map ultrasound measurements to lung aeration maps, providing an efficient and practical surrogate for this complex inverse problem.


\method has certain limitations. 
1) The lack of aligned ex-vivo lung aeration maps prevents full verification of our 2D reconstructions in such setups. Furthermore, the current model has yet to be validated in vivo with human subjects. 2) Lung ultrasound is inherently real-time and interactive, relying on operator adjustments during imaging. \method currently has a runtime of 0.1 seconds, which should be further accelerated for real-time use. 
3) \method has not been linked to final diagnosis, such as ARDS and cardiogenic pulmonary edema (CPE).
Addressing these limitations will allow \method to further advance the interpretability, reproducibility and diagnostic utility of lung ultrasound for acute and chronic lung diseases, paving the way for broader clinical adoption.

\clearpage
 \section{Methods}


\subsection{Ultrasound Data Simulation}
\label{sec:lussim}

In this section, we discuss the forward problem of lung ultrasound: for a specific lung (with acoustic properties such as speed of sound $c_0$ and density $\rho$), we model the ultrasound propagation and simulate the corresponding acoustic pressure/RF data $p$ received at the body surface. The paired lung acoustic properties and received RF data can be used for training \method as a surrogate to solve the inverse problem.

\bmhead{Numerical Simulation of Radio Frequency Data}
For generation of channel data closely matching RF signals received in human body ultrasound scanning, acoustic pressure field simulator Fullwave-2 ~\cite{pinton2021fullwave} was used. This numerical tool is based on the first principles of sound wave propagation in heterogeneous attenuating medium and accounts for its effects such as distributed aberration (wavefront distortion), reverberation (multiple reflections), multiple scattering and refraction (change of wavefront direction at media interface). Fullwave-2 was successfully validated in various diagnostic and therapeutic scenarios, such as \textit{ex vivo} abdominal measurements ~\cite{trahey2017beamforming, bottenus2019impact}, water tank experiments ~\cite{soulioti2021}, \textit{in vivo} human abdominal measurements ~\cite{dahl2012harmonic}, transcranial brain therapy ~\cite{pinton2011attenuation,pinton2012numerical,pinton2012direct} and traumatic brain injury modeling ~\cite{chandrasekaran2024}.

The Fullwave2 model employs a nonlinear full-wave equation defined by the following equation:
\begin{align}
     & \nabla_1 p + \rho \cfrac{\partial \matr{v}}{\partial t} = 0         \label{eq:wave1} \\
     & \nabla_2 \cdot \matr{v} + \kappa \cfrac{\partial p}{\partial t} = 0 \label{eq:wave2}
\end{align}
where $p(\matr{x}, t)$ and $\matr{v}(\matr{x}, t)$ represent the pressure and velocity wavefield at a given position $\matr{x}$ at a given time $t$ respectively,
and $\rho(\matr{x})$ and $\kappa(\matr{x})$ denote the density and the compressibility of the medium at position $\matr{x}$, respectively.
When the nonlinear effects are included,
the density $\rho$ and the compressibility $\kappa$ are modified by the pressure $p$ and the nonlinearity coefficient $\beta=1+B/A$ as follows:
\begin{align}
    \rho   & = \rho_0\left[1 + \kappa_0 p\right]                \\
    \kappa & = \kappa_0\left[1 + \kappa_0 (1 - 2\beta)p)\right]
\end{align}
where $\rho_0$ is the equilibrium density and $\kappa_0$ is the equilibrium compressibility.
Now, $\nabla_1$ and $\nabla_2$ in equations (\ref{eq:wave1}) and (\ref{eq:wave2}) are used to denote the complex spatial derivatives
that model attenuation and dispersion while maintaining
the pressure-velocity formulation of the wave equation.
The complex spatial derivatives $\nabla_1$, $\nabla_2$ are written as a scaling of the partial derivative
and a sum of convolutions with $N$ relaxation functions, which for $\nabla_1$ can be written as
\begin{align}
    \cfrac{\partial}{\partial \tilde{x}_1} & = \cfrac{1}{\kappa_{x_1}}\cfrac{\partial}{\partial x} + \sum^N_{\nu=1}\zeta^\nu_{x_1}(t) \ast\cfrac{\partial}{\partial x} \label{eq:relaxation1} \\
    \cfrac{\partial}{\partial \tilde{y}_1} & = \cfrac{1}{\kappa_{y_1}}\cfrac{\partial}{\partial y} + \sum^N_{\nu=1}\zeta^\nu_{y_1}(t) \ast\cfrac{\partial}{\partial y} \label{eq:relaxation2} \\
    \cfrac{\partial}{\partial \tilde{z}_1} & = \cfrac{1}{\kappa_{z_1}}\cfrac{\partial}{\partial z} + \sum^N_{\nu=1}\zeta^\nu_{z_1}(t) \ast\cfrac{\partial}{\partial z} \label{eq:relaxation3}
\end{align}
$\zeta^\nu_{x_1}$ is the convolution kernel for the $N$ relaxations, indexed by $\nu$. The convolution kernels are defined as:
\begin{align}
    \zeta^\nu_{x_1}(t) & = - \cfrac{d^\nu_{x1}}{\kappa^2_{x_1}} \: e^{-\left(\frac{d^\nu_{x1}}{\kappa_{x_1}} + \alpha^\nu_{x_1}\right)t} H(t) \\
    \zeta^\nu_{y_1}(t) & = - \cfrac{d^\nu_{y1}}{\kappa^2_{y_1}} \: e^{-\left(\frac{d^\nu_{y1}}{\kappa_{y_1}} + \alpha^\nu_{y_1}\right)t} H(t) \\
    \zeta^\nu_{z_1}(t) & = - \cfrac{d^\nu_{z1}}{\kappa^2_{z_1}} \: e^{-\left(\frac{d^\nu_{z1}}{\kappa_{z_1}} + \alpha^\nu_{z_1}\right)t} H(t)
\end{align}
Note that $\kappa_{x_1}(\matr{x})$, $\kappa_{y_1}(\matr{x})$, $\kappa_{z_1}(\matr{x})$ represent a linear scaling of the derivative at position $\matr{x}$.
This scaling parameter modifies the wave velocity in the $x$, $y$ and $z$ directions.
The variables $d^\nu_{x1}$, $d^\nu_{y1}$, $d^\nu_{z1}$, represent a scaling-dependent damping profile;
and $\alpha^\nu_{x_1}$, $\alpha^\nu_{y_1}$, $\alpha^\nu_{z_1}$ denote a scaling-independent damping profile.
$H(t)$ is the Heaviside or unit step function.
The transformation set for the $\nabla_2$ operator
is identical to that of $\nabla_1$ and the variables associated with this second transformation are denoted by the subscript 2.
These relaxation mechanisms incorporated in $\nabla_1$, $\nabla_2$ are introduced to empirically model attenuation based on observations of the attenuation laws and parameters observed in soft tissue. These mechanisms can be generalized to arbitrary attenuation laws through a process of fitting the relaxation constants.

Numerically, Fullwave2 uses the finite-difference time-domain (FDTD) method to solve the nonlinear full-wave equations (\ref{eq:wave1}) and (\ref{eq:wave2}).
In order to ensure numerical stability and accuracy in heterogeneous media with high contrast,
Fullwave2 utilizes the staggered-grid finite difference (FD) discretization, whose FD operator has 2$M$-th order accuracy in space and fourth-order accuracy in time.
\cite{pinton2021fullwave} provides a detailed numerical solution for the computation of the spatial derivatives $\nabla_1$, $\nabla_2$ and relaxation functions.

In the inverse problem setting, the RF data $p_S$ is only received at surface $p_S = Sp$, where $S$ is a linear operator. For simplicity, we refer to the received RF data as $p$ in the paper. Density map $\rho$ contains two parts: chest wall and lung. Lung aeration map $\rho^A$ is only the lung segmentation from $\rho$, which is flattened and thresholded to a binary map, with 1 denoting air pixels and 0 denoting non-air (tissue) pixels (details in the later part of this subsection).

Key acoustical properties of the air, such as almost total reflectivity and impermeability due to high impedance mismatch with soft tissue, were modeled via constant zero pressure in air inclusions similar to~\cite{ostras2021,ostras2023}. This approach is beneficial in terms of computational cost and simulation stability compared to pulse propagation in medium of low sound speed and mass density.

A clinically relevant scenario of use of linear transducer L12-5 50~mm (ATL, Bothell, Washington, USA) and 128 sequential focused transmit (2-cycle 2.5~MPa pulses, walking 64 element aperture) at center frequency 5.2~MHz was simulated~\cite{demi2022,rocca2023}. This transducer is compatible with research ultrasound systems, such as Vantage 256 (Verasonics Inc, Kirkland, Washington, USA) and a list of diagnostic ultrasound imaging systems Philips. The focal depth varied from 1 to 3~cm and was set to the lung surface position (pleura depth) to maintain the maximum energy deposition at the acoustical channels entrance and for optimal visualization of diagnostic features - vertical artifacts~\cite{mento2020,mento2021}. Accordingly, transmit f-number varied from 0.8 to 2.4. The RF signals were received at locations of the same fired 64 transducer elements (subaperture) and sampled at a rate of 20.8~MHz with 70\% fractional bandwidth which is comparable to modern clinical ultrasound imaging systems. All the simulations were performed in 2D space, each independent transmit-receive event in a field of 2.5~cm width and 4.5~cm depth with 0.195~mm lateral translation between neighbor subapertures/events. The duration of the simulations was limited to 87.6~$\mu$s for compliance with existing clinical recommendations~\cite{dietrich2016}, to be able to visualize at least the second reflection of the ballistic pulse between surfaces of transducers and the lung-horizontal artifact in healthy cases~\cite{soldati2020}. A spatial grid of 12 points per wavelength (24.6~$\mu$m) and temporal discretization was set to 8.0~ns accounting for a reference speed sound of 1540~m/s, which corresponded to a Courant-Friedrichs-Lewy condition of 0.5~\cite{courant1967}.

\bmhead{Acoustical Properties of the Human Body Wall and the Lung} 
Maps of acoustical properties were combined out of segmented axial anatomical images of human body wall (Visible Human Project ~\cite{spitzer1996}, resolution 330~$\mu$m) and histological images (hematoxylin and eosin staining, section thickness 5~$\mu$m) of healthy swine lung, resolution 0.55~$\mu$m (c\ref{fig:data}a-c), similar to ~\cite{ostras2023}. Segmented slides from these sources were interpolated (nearest neighbor) to fit the simulation elements size (24.6~$\mu$m). Fresh tissue for histological processing and \textit{ex vivo} ultrasound scanning was provided by Tissue Sharing Program of North Carolina State University and processed by Pathology Services Core, University of North Carolina at Chapel Hill. Slide scanner SlideView~\textsuperscript{TM} VS200 (Olympus Corporation, Tokyo, Japan) was used for bright-field microscopy at 10x magnification. Photographic images of human body cryosections were segmented into three types of tissue (connective, adipose and muscle) using custom tissue-specific probability distribution functions in 3 color channels (RGB). Parietal and visceral pleura borders were segmented manually in the original cryosection and histological images. It allowed axial column translation for flattening of the lung and its deformation to conform with various parietal pleura curvature. Histological lung images were binary segmented (air / non-air) based on the arbitrarily chosen brightness threshold of 0.8 in the green channel.

A wide variety of alveolar size (median 94 [IQR 72-132]~$\mu$m) and alveolar wall thickness (median 16.5 [IQR 5.5-38.5]~$\mu$m) was assured by use of lung tissue from 10 animals of different age (median 3.5 [IQR 3-6] months) and weight (median 78 [IQR 60-185]~kg). While the lateral position of vertical artifacts in B-mode images correlates with the location of acoustical channels ~\cite{kameda2022,ostras2023}, transfer of the superficial layer (200~$\mu$m) of the histological part of the map allowed significant increase variability of the observed artifacts and underlying RF data (\cref{fig:sup_data}). Leveled and cropped rectangular lung tissue images of size 1.8x1.2~cm were randomly selected from the 521 unit dataset and tailed to obtain a continuous 5~cm wide lung layer for each simulated image. Both flattened (X cases) and naturally curved (Y cases) lung layers were employed. To cover a range of diagnostic ultrasound image features (horizontal and vertical artifacts), values of aeration and depth of the lung parenchyma were randomly selected out of uniform distributions in the range [10 90] \% and [1 3]~cm, respectively. 
Lung aeration was calculated as a percent of air elements out of all lung parenchymal elements in 2D. Alveolar size and alveolar wall thickness were calculated as linear intercepts based on standard method~\cite{knudsen2010} using custom Matlab (Mathworks, Natick, MA, USA) code.
Uniform alveolar derecruitment characteristic for Acute Respiratory Distress Syndrome (ARDS) at the air-alveoli interface was modeled numerically. First, the subpopulation of air pixels \(A\) neighboring with non-air ones was found. Second, the number of air pixels \(n\) necessary to convert into non-air to achieve target aeration was calculated. Finally, \(n\) random pixels were drawn out of subpopulation \(A\) and converted into non-air. In case if \(n\) was more than the number of pixels in \(A\), all the subpopulation was converted and the algorithm was repeated while the target aeration is achieved. If target aeration was higher than the initial one, the same sequence was performed finding non-air pixels in step 1 and converting to air in step 3. These alterations facilitated practically continuous variable aeration.

\bmhead{Beamforming}
We introduce the beamforming used in the paper as a baseline to transform RF data into conventional human-readable B-mode images.
Using a conventional delay-and-sum (DAS) beamforming algorithm, a dataset of delayed RF signals and corresponding reconstructed 2.5~cm wide B-mode images were composed out of both numerically simulated and scanned \textit{ex vivo} data. Each tensor in the form [time, receiver, transmit-receive event] represented a single scan/image consisting of signals from 64 receivers in 128 independent events. Individual receivers' vector signals $s_m$ were delayed by precalculated $\tau_m$ samples to ensure proper focusing, compensate differences in their (receivers) spatial/lateral position and synchronize these readings ~\cite{mckeighen1977} assuming a homogeneous speed of sound 1540~m/s. The beamformed signal of a single transmit-receive event can be defined in the discrete-time domain as:

\begin{equation}
\label{eq:3}
bfs(t) = \sum_{m=1}^{M} {s_m}(t-{\tau_m})                        
\end{equation}

Where \(bfs(t)\) is output signal, \(m\) - number of the receiver (1..64), \(t\) - discrete time.

DAS has multiple advantages, such as i) grounding on the basic principles of wave propagation; ii) simple implementation; iii) low computational cost and possible parallelization, which makes it applicable in real-time; and iv) statistics of real envelopes ~\cite{destrempes2010} and temporal coherence ~\cite{salles2014} are preserved. Numerical robustness and data-independency make this beamforming technique highly generalizable and the most widely used not only in ultrasound imaging, but also in telecommunication.

After DAS beamforming, RF signals were envelope-detected using Hilbert transform and log compressed to a dynamic range appropriate for human visualization [-60 0] ~dB. Such processing is conventional and allows to emphasize on weak scattering along with high-intensity reflections using the same scale ~\cite{park2023}.   

For smoother visual representation, isotropic 2D interpolation (bicubic, factor 4) was applied to B-mode images. To preserve the generalizability of acquired RF data, TGC was not applied to them ~\cite{vara2020}. The image processing was performed in Matlab (Mathworks, Natick, MA, USA). The numerical simulations were done on a Linux-based computer cluster running GPU NVIDIA\textsuperscript{\textregistered} V-100. Individual simulations (128 per B-mode image) were running in parallel and took up to 6 hours.


\subsection{Ex Vivo Data}
\label{sec:realdata}
\bmhead{Ex-Vivo Data Acquisition}

Fresh tissues of 12 animals (median age 3.5 [IQR 1.75-6] months), median weight 85 [IQR 50-180]~kg) were hand-held scanned in a water tank using linear transducer L12-5 50~mm (ATL, Bothell, Washington, USA) and research ultrasound system Vantage 256 (Verasonics Inc, Kirkland, Washington, USA). The sequence, time discretization and bandwidth parameters were identical to those described in simulations. The focal depth varied from X to Y~cm and was manually set by the operator corresponding to the visualized pleural depth. Time gain compensation (TGC) was applied and individually adjusted for proper visualization of relevant diagnostic features in the scanning process, however, RF data were saved and analyzed without compensation for generalizability reasons. 124 scans were performed (118 for assessment of network performance and 6 for calibration of the simulations).
To reproduce intercostal views of transthoracic lung ultrasound imaging, two tissue layers were placed in the water tank - body wall (chest or abdominal) at the top and lung on the bottom. The former was completely submerged in degassed and deionized water and fixed/immobilized to the water tank walls with a custom plastic fixture and nylon strings sutured through the tissue (outside the field of view). The latter was placed costal surface toward the body wall and transducer in anatomical position and free-floating. The bronchus was intubated with custom plastic fitting hermetically connected to i) manual rubber bulb air pump with bleed valve, ii) airway manometer (leveled water column pressure gauge) and iii) line with 3-way stopcock and syringe port outside the water tank for fluid instillation in airways. Pulmonary arteries were ligated and sutured twice and veins were sutured twice at harvesting to prevent aeration of lung vasculature.
Lobes of left swine lungs were scanned due to their closest gross anatomical similarity to human ones~\cite{judge2014}. Both lungs were dissected into lobes (two in the left, three in the right lung) and segments when technically possible using intersegmental veins and inflation-deflation lines as guides~\cite{oizumi2014}. The dissected lobes/segments were tested for aerostasis (hermeticity) and units with air leaks were excluded from further study (success rate 48\%). 

\bmhead{Aeration Calculation} Units obtained out of the right lung and caudal lobe of the left lung were employed in the calculation of its bulk aeration, while the left cranial lobe was used for scanning. Such split was necessary because of the destructive nature of degassing after which lung tissue is altered and does not represent normal anatomy at micro- and mesoscopic levels ~\cite{scarpelli1998}. The lung lobe was short-term inflated to an airway pressure of 20 cm H\textsubscript{2}O for alveolar recruitment and static continuous positive airway pressure of 10 cm H\textsubscript{2}O ~\cite{braithwaite2023,lee2009} was maintained during volume measurements and scanning. Weighing was performed with the bronchus cross-clamped and the airway fitting disconnected. Bulk lung aeration was calculated as \(A_l = (V_l-V_t)/V_l\), where \(V_l\) is the total volume of the lobe measured using the fluid displacement method and \(V_t\) - volume of the lobe after de-aeration. Lung tissue was degassed three times~\cite{stengel1980} for 10 minutes in a vacuum chamber at pressure -25 mm Hg~\cite{edmunds1967}. In addition to aeration, lung mass density was calculated as \(\rho_l = m_l/V_l\), where \(m_l\)  - total weight of the lobe.
To model ARDS-like distributed deaeration airway/bronchial instillation of isotonic NaCl solution was employed. The severity of lesion was leveraged by adding different volumes of the fluid.

\bmhead{Summary: Data used in the Study} Our machine learning (ML) framework \method is mainly trained on simulated data and evaluated on real ex-vivo data.
We summarize the data used in the paper in Table \ref{tab:data_comp}.

\begin{figure}[t!]
    \centering
   \includegraphics[width=\columnwidth]{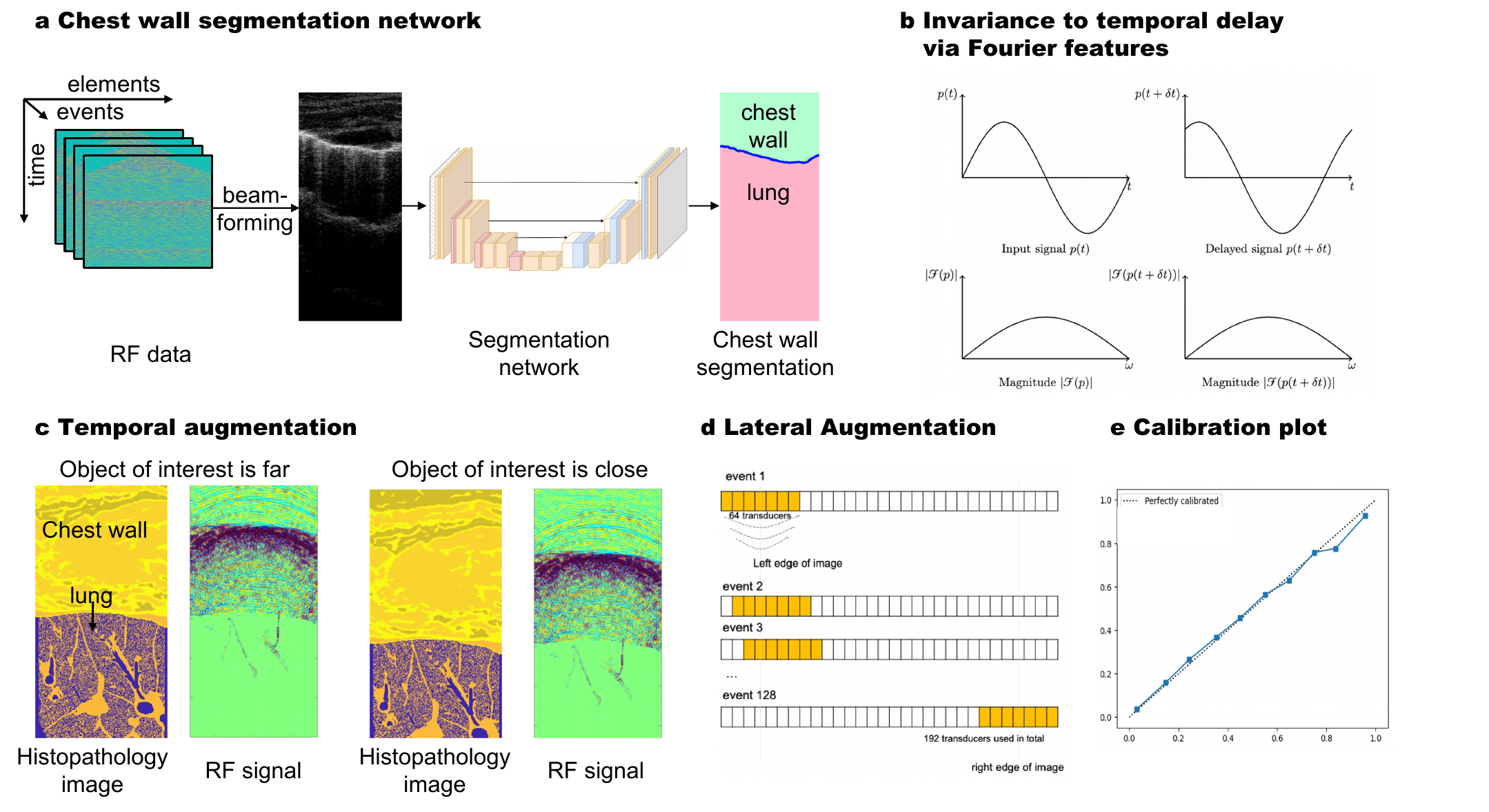} 
   \caption{ \small
{\bf Architecture design and training details of \method.}  
{\bf a,} Chest wall segmentation network: This network reconstructs the chest wall segmentation from the RF data.  
{\bf b,} Fourier feature learning allows invariance
learning to the temporal delay (with details in Section \ref{sec:architect}). 
{\bf c,} Temporal augmentation: The temporal dimension ($t=0$ to $t=200$) is randomly masked during training, excluding the initial signal corresponding to chest wall structures, which are irrelevant to aeration reconstruction.  
{\bf d,} Frequency-space processing: \method operates directly in the frequency domain, where Fourier features provide invariance to temporal delays in the RF data and variations in chest wall depth.  
{\bf e,} Calibration plot: \method demonstrates strong calibration, with predicted aeration percentages closely matching true values, ensuring accurate and reliable predictions.
}
    \label{fig:mlmethod}
\end{figure}

\subsection{\method Architecture}
\label{sec:architect}


\bmhead{Overview} \method reconstructs lung aeration map $\rho^A$ from the measured RF data $p$.
The overall \method pipeline is in \cref{fig:data}a. 
The RF signal is three-dimensional, denoted as $p \in \mathbb{R}^{ T\times N_{t} \times N_{e} }$, where $T, N_{t}, N_e$ refer to the number of temporal steps, the number of transducer elements and the number of events, the distinct ultrasound pulse emissions in a scan. Before the reconstruction of $\rho^A$, we use a segmentation model $\mathcal{S}$ to estimate the chest wall segmentation map $\mathcal{S}(p)$ as an auxiliary task to inform the reconstruction model of the chest wall structure.
We then combine the predicted chest wall segmentation map with RF data and predict the lung aeration map $\hat{\rho}^A$ with a neural operator $f_{\theta}$:
\begin{align}
    \hat{\rho}^A = f_{\theta}(p,\mathcal{S}(p))
\end{align}
where $\theta$ are the parameters of the neural operator.
To train the neural operator, we use a loss $\mathcal{L}$ to penalize the difference between predicted lung aeration map $\hat{\rho}^A$ and the ground truth $\rho^A$, i.e. $\min_{\theta} \mathcal{L}(\hat{\rho^A}, \rho^A)$.
Next, we provide more details of each module of \method, followed by the simulation to real domain adaption and the model calibration. Ablation study that demonstrates the empircal contribution of different modules and design choices are in the Section \ref{sec:ablation} of the Supplementary. 


\bmhead{Chest Wall Segmentation} 
Considering the spatial correspondence between the B-mode image and the chest wall segmentation map (\cref{fig:mlmethod}a), we first beamform the lung ultrasound RF data $p$ to B-mode images $\mathcal{B}(p)$ and then feed the B-mode images to the segmentation model (see beamforming details in Section \ref{sec:lussim}).
During training, the B-mode images $\mathcal{B}(p)$ and ground-truth lung chest wall segmentation map $S$ are resized to $400\times400$ for the efficient training of the network.
The segmentation network adopts a UNet \cite{ronneberger2015u} architecture, with 4 downsampling layers with stride 2. 
The downsampling path reduces the resized B-mode images $\mathcal{B}(p)$  down to \(25 \times 25\) at the bottleneck. The downsampling path also increases the feature channel from $1$ to $64, 128, 256, 512$ sequentially. 
The upsampling path then reconstructs the spatial dimensions, combining features from the corresponding downsampling layers to preserve spatial information. Finally, an output convolutional layer produces the segmentation map $\mathcal{S}(p)$ of size $400 \times 400 \times 2$, where the first channel corresponds to the chest wall region and the second one to the lung region. 
To train the network, we penalize the difference between the prediction and the ground truth with a cross-entropy loss $\mathcal{L}_{\text{CE-seg}}(\mathcal{S}(p), S)$.

On the in silico test set, our model achieves  95.8\% under the Dice similarity coefficient, a well-accepted metric for evaluating image segmentation performance \cite{dice1945measures}.   Additional results of the auxiliary task, chest wall segmentation, can be found in the Supplementary (Section \ref{sec:seg_results}). 

\bmhead{Aeration Map Reconstruction} The measured lung ultrasound RF data $p$ is combined with chest wall segmentation $\mathcal{S}(p)$ to reconstruct the lung aeration map $\rho^A$.
The reconstruction network is designed as a temporal neural operator which enables invariance learning to the temporal delay (\cref{fig:mlmethod}b), followed by a spatial network to capture the local features for the aeration map reconstruction. To improve the robustness of the model, we adopt a data augmentation strategy involving both depth and lateral masking (\cref{fig:mlmethod}c,d). 

\underline{\it Background: Fourier Neural Operator}. FNO (Fourier neural operator) \cite{li2021fourier} is a powerful neural operator framework that efficiently learns mappings in function spaces, with many applications as surrogate models for solving partial differential equations (PDEs)  with many applications \cite{rashid2022learning,pathak2022fourcastnet,guan2023fourier}. In our approach, we decompose 3D ultrasound RF data into temporal and transducer dimensions (element$\times$event) to achieve computational efficiency by handling the high-dimensional data in a structured way. Applying FNO specifically to the temporal dimension leverages FNO’s strengths in capturing time-dependent patterns, enabling efficient and precise modeling of the data while preserving computational speed and accuracy.

 \underline{\it Fourier Neural Operator on Temporal Dimension for Invariance to Temporal Delay}.
Our input data $p$'s first dimension is a temporal signal and the output is invariant to delay, i.e., $p(t)$ and $p(t+\delta t)$ are mapped to the same output. In our network, we propose to use the Fourier transform to get the Fourier feature $\mathcal{F}(p)$.
Specifically, $p$ has a size of $T \times N_t \times N_e $, where $T, N_e, N_t$ represent the temporal dimension,  the number of transducer elements and the the number of events.  We perform a Fourier transform on the temporal dimension \( T \) only, treating the other dimensions \( N_t \) and \( N_e \) as batches. The Fourier transform of \( p \) along the \( T \) dimension can be denoted as $\mathcal{p}(f, n_e, n_t) = \mathcal{F}\{p(t, n_t, n_e)\}$, where \( \mathcal{F} \) denotes the Fourier transform on the temporal domain, \( f \) is the frequency domain corresponding to the temporal dimension. \( T \), \( n_e \) is the index in the \( N_e \) dimension and \( n_t \) is the index in the \( N_t \) dimension.
The Fourier transformed data \( P \) can be represented in terms of its magnitude and phase as follows:
\begin{align}
    |\mathcal{p}(f, n_e, n_t)| &= \text{Magnitude of } \mathcal{p}(f, n_e, n_t) \\
    \angle \mathcal{p}(f, n_e, n_t) &= \text{Phase of } \mathcal{p}(f, n_e, n_t)
\end{align}
Features are then learned from both the magnitude and phase of the Fourier-transformed data. Because the magnitude $|\mathcal{F}(p)|$ is linear time-invariant(invariant to delay), it can learn the delay-invariance mapping effectively.

The FNO is implemented with a one-dimensional process on the temporal dimension (by treating other dimensions as different samples of a batch). 
The FNO module consists of 2 layers, each of which has 32 output channels and 87 modes in the Fourier domain.

\underline{\it Spatial Network on Transducer Element Dimensions for Lateral Reconstruction}.
We use a spatial network consisting of a series of convolutions after the FNO module for processing the RF data's transducer element dimensions (element $n_t\times$event $n_e$). 
The spatial network also allows learning local features, which is important for image reconstruction and augments FNO's global feature learning, as FNO's Fourier space feature learning can be considered as applying global convolution to the RF signal. 
The network consists of a series of downsampling and upsampling layers, with a similar design to \cite{ronneberger2015u}. The downsampling path includes layers with channel transitions of 64 to 64, 64 to 128, 128 to 256 and 256 to 256 (due to bilinear interpolation). The upsampling path includes layers with channel transitions of 512 to 256, 256 to 128, 128 to 64, 96 to 32 and a final transposed convolutional layer from 32 to 32, followed by a batch normalization layer for 32 channels.

\underline{\it Masking-Based Data Augmentation}.
To increase the robustness of \method and to make the aeration map $\rho^A$ reconstruction invariant to partial masking $m$ of the RF signal $p$, we adopt local-masking-based data augmentation strategies during training. That is, the masked input $m(p)$ and unmasked input $p$ should map to the same reconstruction. 
During training, we consider two different masking strategies:
\begin{enumerate}
    \item {\bf Temporal Masking in RF for depth dimension of the reconstruction}. The goal is to strengthen the invariance to temporal delay and chest wall depth from FNO's design via additional data augmentation. The temporal masking is adopted to enable other layers to also be invariant. Specifically, we uniformly masked the initial $M_t$ steps ($M_t \leq 200$, or 13\% of the total temporal length). The reason is that the initial temporal steps correspond to the reflections in the chest wall, to which we want to make the reconstruction invariant.
    \item {\bf Transducer Element Masking in RF for the lateral dimension of the reconstruction}. To encourage lateral correspondence, we randomly mask $M_s$ elements of the flattened transducer element dimension of size $N_e \cdot N_t$.  $M_s \leq 2000$, or 24\% of the total number of equivalent transducer size.
\end{enumerate}


\bmhead{Loss Functions}
We optimize \method by minimizing the training loss, which consists of the 2D aeration map reconstruction loss and aeration prediction loss.
The  2D aeration map reconstruction loss $\mathcal{L}_{\text{CE}}$ is a cross-entropy loss that measures the difference between the predicted and ground-truth aeration maps:
\begin{equation}
\mathcal{L}_{\text{CE}} = -\sum_{i,j} \left[ \rho_{ij} \log \hat{\rho}_{ij} + (1 - \rho_{ij}) \log (1 - \hat{\rho}_{ij}) \right]
\end{equation}
where $\rho_{ij}$ is the $i^{\text{th}}, j^{\text{th}}$ pixel of ground-truth aeration map $\rho^A$ and $\hat{\rho}_{ij}$ is the $i^{\text{th}}, j^{\text{th}}$ pixel of reconstructed aeration map $\hat{\rho}^A$

The aeration prediction loss is a $\ell_1$ loss that penalizes the difference between the predicted aeration $\hat{\gamma}$ and ground-truth $\gamma$.
\begin{align}
  \mathcal{L}_{\gamma} = \| \gamma - \hat{\gamma}\|
\end{align}
The total loss is a combination of 2D aeration map reconstruction loss and aeration prediction loss:
\begin{align}
\label{eq:comb_loss}
    \mathcal{L} = \mathcal{L}_{\text{CE}}+ \eta \mathcal{L}_{\gamma}
\end{align}
where $\eta$ is the weight of the aeration loss, which we set to be 0.5 in the experiments.

\bmhead{Simulation to Real Data Adaption}
White Gaussian noise is added to simulated RF data during training on the simulated data to reduce the gap between simulated and real data. 
After finishing training the model with simulated data, we fine-tune with $18$ real data samples, with details available in Table \ref{tab:data_comp}. During fine-tuning, we only use aeration loss $\mathcal{L}_{\gamma}$ as the 2D aeration map of the real samples is not available.

\bmhead{Model Calibration}
Model calibration measures how well the model’s confidence in its predictions matches its actual accuracy \cite{guo2017calibration}. Ideally, a well-calibrated model has confidence levels proportional to its prediction accuracy. We conduct a calibration test of \method for reconstructing aeration maps, showing that it performs well in simulation (\cref{fig:mlmethod}e).  

To refine confidence levels, we applied a simple scaling method called Platt scaling \cite{platt1999probabilistic}, which adjusts predictions using a sigmoid function: $\tilde{\rho}^A_{ij} = \sigma(\hat{\rho}^A_{ij})$, where $\sigma(x) = \frac{1}{1 + e^{-x}}$. This ensures predictions better reflect true aeration percentages.

The calibration results are reported in \cref{fig:mlmethod}e. We observe that the model is well-calibrated in silicon. Note that the calibration is not available on the real data as the ground-truth 2D aeration maps are unavailable. 
The calibration techniques help in adjusting the confidence scores of \method's predictions to better reflect their true accuracy, ensuring that the confidence level is a reliable indicator of the model's performance.

\subsection{Implementation Details and Evaluation Protocols}
\label{sec:implemntation}
\bmhead{Fullwave-2 Solver for Simulated lung ultrasound Data Generation}
We simulate the wave propagation in the lung with the finite difference method with nonlinear wave propagation \cite{pinton2021fullwave}. With a batch size of 300, the solver takes approximately 36 hours on 16 NVIDIA V100 GPUs.

\bmhead{Chest Wall Segmentation Model} We use RMSprop optimizer \cite{rmsprop} with a learning rate of $ 1\times10^{-5}$, weight decay of $ 1\times10^{-8}$ and momentum of $0.999$. The batch size is 5. We train the model for 10 epochs on one NVIDIA 4090 GPU.
\bmhead{Aeration Reconstruction Model}
We use Adam optimizer \cite{kingma2014adam} with a learning rate of $\eta \cdot 0.002$ and $\beta = (0, 0.99^{\eta})$. 
 The model is implemented with the PyTorch framework. The batch size is 26. We first train the model with simulated data for 90 epochs. We then fine-tune the model with real data for another 10 epochs. In total, our training took 1.5 days on one NVIDIA A100 GPU.

\bmhead{Evaluation Protocols}
We adopt several metrics to evaluate the performance of \method.
The first one is the error of the predicted lung aeration. 
\begin{align}
    \text{aeration error} = 
    \| \gamma-\hat{\gamma}\|
\end{align}
where $\gamma, \hat{\gamma}$ refers to the aeration of ground-truth and predicted lung aeration map $\rho^A, \hat{\rho}^A$.
For simulation data, $\rho^A$ can be calculated according to Eqn. \ref{eq:aeration}. For real data, the aeration is measured with weighting, with details in Section \ref{sec:realdata}. 


As the ground-truth lung aeration map is also available for simulation,
we consider two other metrics to evaluate the 2D reconstruction quality.
\begin{enumerate}
    \item The Normalized Mean Squared Error (NMSE) measures the average of the squares of the errors normalized by the ground truth image's energy:
$\text{NMSE} = \frac{\sum_{i,j}^{H,W} \left( \rho^A_{i,j} 
- \hat{\rho}^A_{i,j} \right)^2}
{\sum_{i,j}^{H,W} (\rho^A_{i,j}) ^2}$. 
\item The Peak Signal-to-Noise Ratio (PSNR) measures the ratio between the maximum possible power of a signal and the power of corrupting noise that affects the fidelity of its representation:
$\text{PSNR} = 10 \cdot \log_{10} \left( \frac{\max(\rho^A_{i,j})^2}{\frac{1}{HW} \sum_{i,j} \left( \rho^A_{i,j}  - \hat{\rho}^A_{i,j}  \right)^2} \right)$.
Note that $H$ and $W$ are the height and width of the image, respectively.
\end{enumerate}




\subsection*{Acknowledgments}
 This work is supported in part by NIH-R21EB033150
 and ONR (MURI grant N000142312654 and N000142012786). A.A. is supported in part by Bren endowed chair and the AI2050 senior fellow program at Schmidt Sciences. B. T. is supported in part by the Swartz Foundation Fellowship. Z. Li is supported in part by NVIDIA Fellowship. The authors thank Zezhou Cheng and Julius Berner for their helpful discussions. 

\subsection*{Competing Interests}
The authors declare no competing interests.
\clearpage

\bibliography{references.bib}

\newpage
\section*{Supplementary Information}
\begin{appendices}
 
\section{Data Generation and Acquisition Details}
\label{sec:hist}

This section describes the creation of lung histology maps for simulating lung ultrasound data. \method is trained on abundant simulation data. 

\begin{figure}[htb]
\vspace{-1.5em}
    \centering
   \includegraphics[width=0.9\columnwidth]{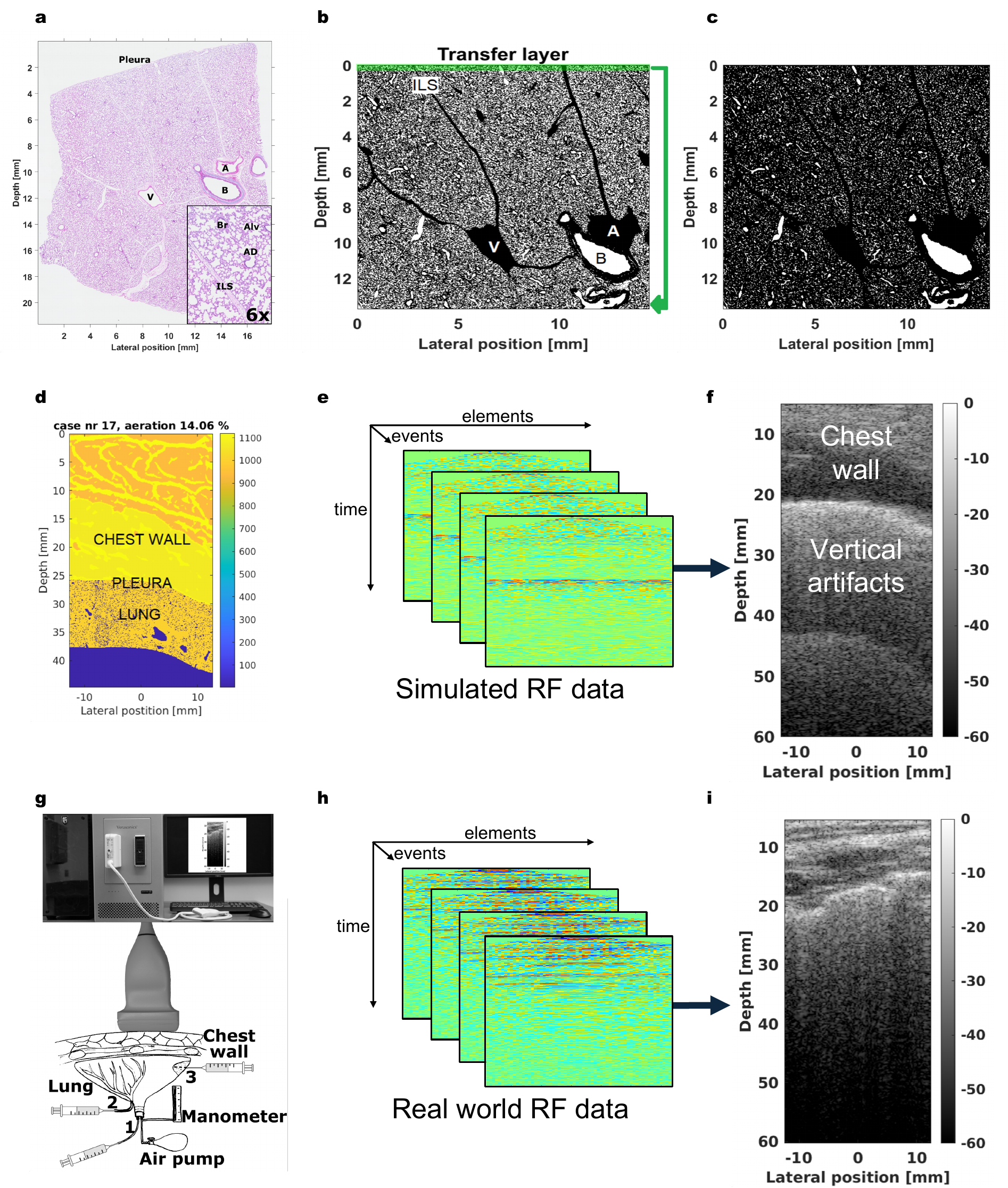}
   \caption{ \small {\bf The process of ultrasound imaging \textit{in silico} (a-f) and \textit{ex vivo} (g-i).} \method is trained on 10k in silico (simulated) data and fine-tuned on 18 ex vivo (real) data.
    {\bf a,} Formation of the acoustical maps as a stationary input for Fullwave-2 simulation tool: histology (H\&E) of healthy swine lung. Insertion shows a magnified part of it and the ability to identify lung tissue architecture. B - bronchus, Br - bronchiole, A - artery, V - vein, AD - alveolar duct, Alv - alveolus, ILS - interlobular septum.
    {\bf b,} Binary-segmented and leveled histology image deformed to conform to linear pleural line and cropped to a rectangular shape. Its superficial 0.1~mm thick layer (\textcolor{green}{green}) is used for repetitive transfer procedures to increase the spatial variability of lung structures.
    {\bf c,} Segmented histology after applied algorithmic modification modeling ARDS (added fluid/non-air pixels are evenly distributed among tissue-air interfaces) with target aeration of 14\%.
    {\bf d,} Combined aeration map comprised of the tissue-specifically segmented body wall (top) and underlying lung deformed to conform its internal surface which models realistic pleural interface.
    {\bf e,} Stack of numerically simulated raw RF data of 128 transmit-receive events visualized as the amplitude of received backscattered signal in form receiver-time. 
    {\bf f,} Corresponding B-mode image formed using simulated RF data demonstrates its proper anatomical part (chest wall thickness and composition) along with multiple coalescent vertical artifacts below the pleural line which are consistent with modeled uniformly distributed fluid retention in lung parenchyma.
    {\bf g,} Scanning of fresh porcine lungs of known aeration (displacement method) through chest wall fragment in the water tank (\textit{ex vivo}) using a programmable ultrasound machine and linear transducer.
    {\bf h,} Stack of real-world raw RF data of 128 transmit-receive events.
    {\bf i,} B-mode image formed using real-world RF data demonstrates both anatomical (chest wall tissue, pleural line) and artifactual (multiple coalescent vertical artifacts below the pleural line) parts of the image consistent with modeled significant uniform fluid retention in the lung. 
}
    \label{fig:sup_hist}
\end{figure}

Acoustical property maps were generated by combining human chest wall anatomical images (Visible Human Project, 330 $\mu$m resolution) with high-resolution histological images of healthy swine lung tissue (5 $\mu$m thick, 0.55 $\mu$m resolution), following \cite{ostras2023}. Binary-segmented histological images quantified lung aeration (air vs. non-air), producing maps where 1 represents air and 0 represents non-air ($\rho^A: \mathbb{Z}^2 \to {0, 1}$). Details of this process are shown in \cref{fig:sup_hist} along with the simulation data generation and real data acquisition details. 

\begin{figure}[htb]
    \centering
   \includegraphics[width=\columnwidth]{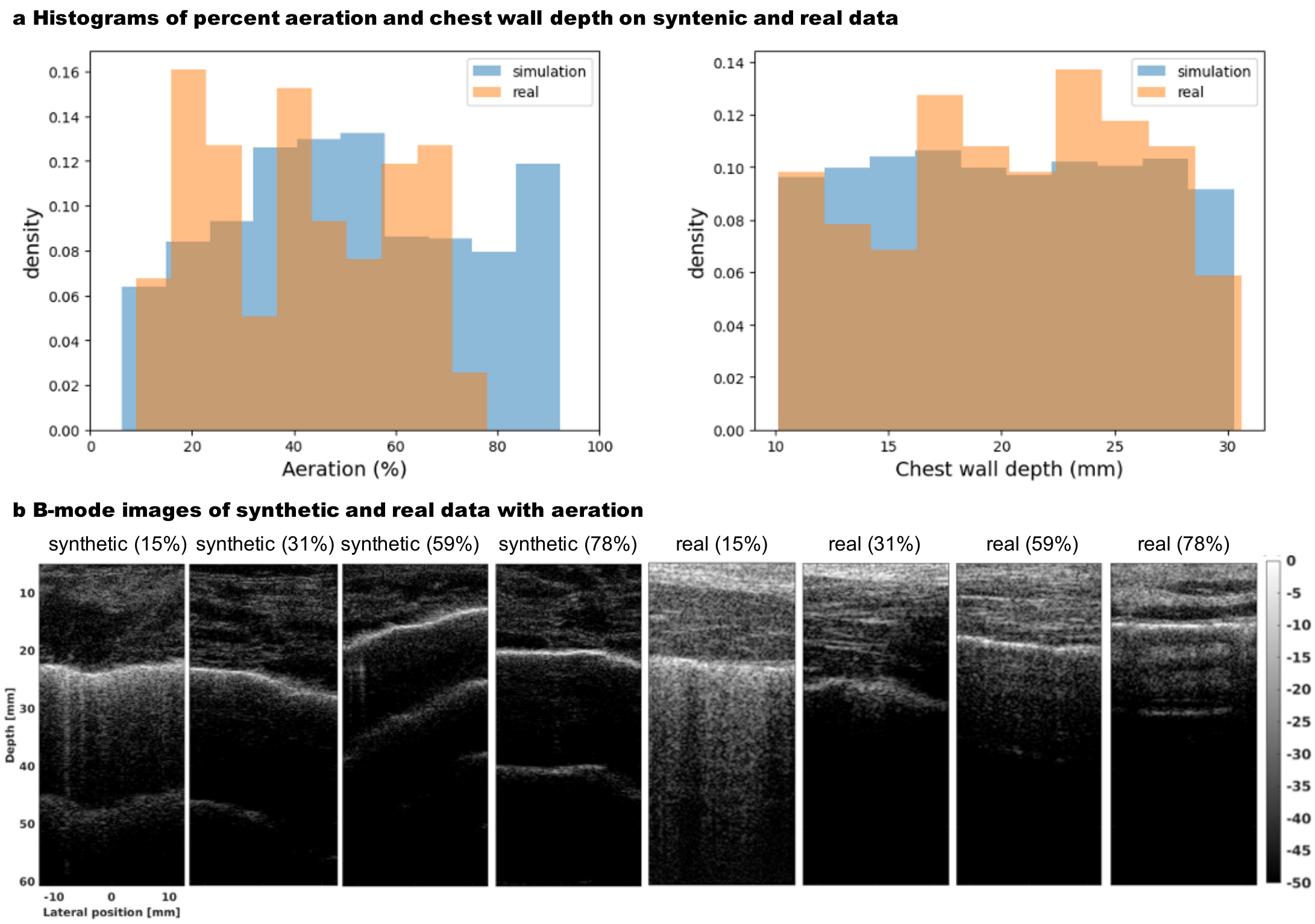}
   \caption{ \small {\bf The distribution and visualizations of the synthetic and real data used to train and fine-tune \method.}
     {\bf a,} The histogram of percent aeration and chest wall depth, two important lung properties, of \textit{in silico} and \textit{ex vivo} data. 
   \textit{In silico} data covers all possible lung properties of \textit{ex vivo} data.
      {\bf b,} B-mode images corresponding to varying levels of aeration used in \textit{in silico} and \textit{ex vivo} experiments, demonstrating a small domain discrepancy between the two datasets.
}
    \label{fig:sup_data}
\end{figure}

\begin{figure}[htb]
    \centering
   \includegraphics[width=\columnwidth]{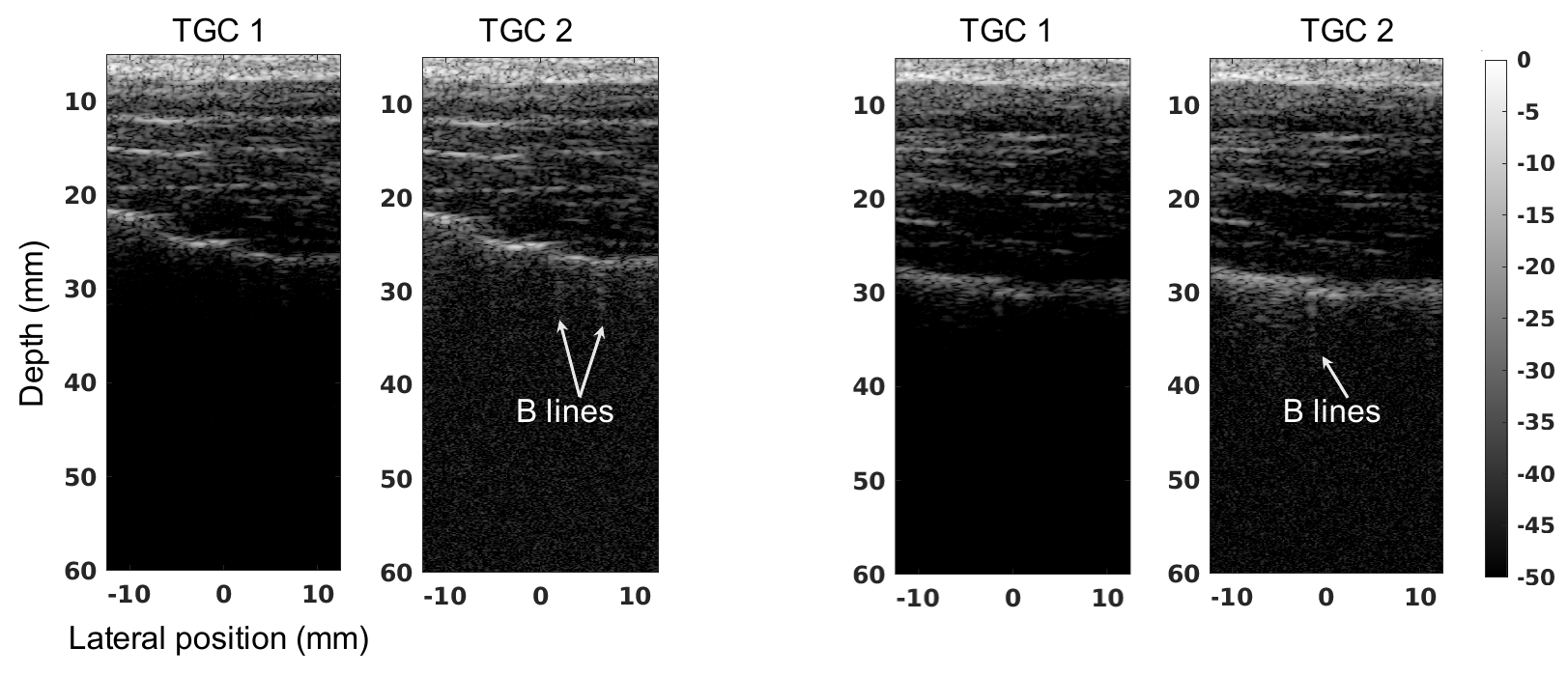}
   \caption{ \small {\bf Different beamforming parameters increase the variation and reduce the reliability of interpreting B-mode images.} 
   Time gain compensation (TGC), a beamforming parameter, changes the B-mode image artifacts and increases the variability for interpreting the lung status.  We show the same sample with two different TGC settings, where under the latter settings artifacts like B-lines are visible. The appearance of B-lines confuses human radiologists and machine learning models that take B-mode images as input, as such artifacts are indicators for low aeration. 
   The left sample has 39\% aeration and the right sample has 29\% aeration.
}
    \label{fig:tgc}
\end{figure}

In \cref{fig:sup_data}, we provide the distribution and visualizations of the synthetic and real data used to train and fine-tune \method. The small gap between the two data set allows the robustness of \method's performance on both synthetic and real data.
In \cref{fig:tgc}, we demonstrate that beamforming parameters increase the variation and reduce the reliability of interpreting B-mode images. 
Thus, machine learning models that take B-mode images as input \cite{horry2020covid,born2020pocovid,roy2020deep}
may not generalize well to different ultrasound devices as each device has a setting with different imaging parameters (dynamic range, time gain compensation).

\section{Chest Wall Segmentation Results}
\label{sec:seg_results}

\begin{figure}[t!]
    \centering
   \includegraphics[width=\columnwidth]{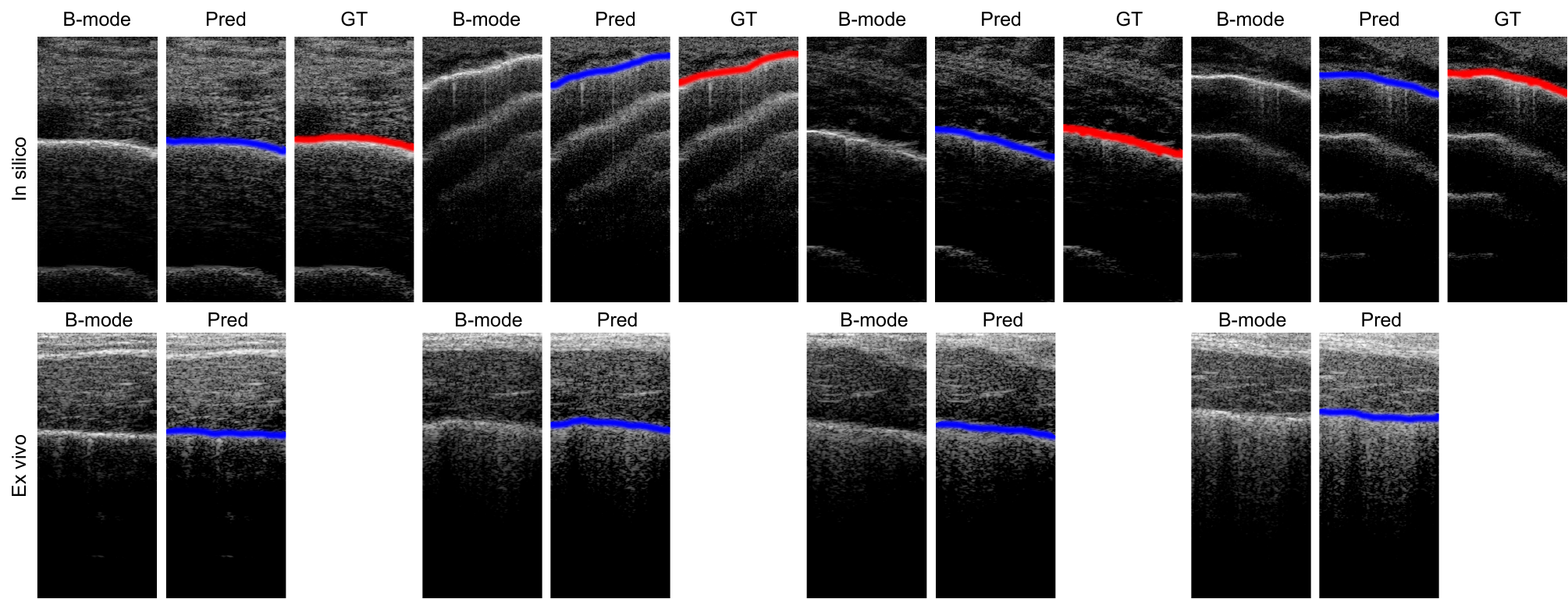}
   \caption{ \small
{\bf Chest wall segmentation results.} The separation line (pleural line) between the chest wall and the lung is overlaid onto B-mode ultrasound images for visualization. Upper: For in silico evaluation samples, the predicted segmentation (Pred) aligns closely with the ground truth (GT) obtained from the simulation. Lower: For ex vivo evaluation samples, the segmentation performance remains robust, with a small performance gap observed between the simulated and real-world datasets.
}
    \label{fig:supp_seg}
\end{figure}

This section presents the results of chest wall segmentation, highlighting the separation line (pleural line) between the chest wall and the lung, which is consistently located below the chest wall in the input B-mode ultrasound images (\cref{fig:supp_seg}). For visualization, the separation line is overlaid onto the B-mode images to demonstrate the segmentation performance. On the in silico test set, our model achieves an impressive Dice similarity coefficient \cite{dice1945measures} of 95.8\%, indicating near-perfect alignment with the ground truth segmentation obtained from the simulation. In the ex vivo test set, the model performs reliably despite the inherent domain differences between simulation and real-world data. The results suggest a minimal performance gap between the in silico and ex vivo conditions, demonstrating the robustness of our method.

\section{Ablation Study on Network Design}
\label{sec:ablation}
\begin{figure}[htb]
    \centering
    \includegraphics[width=0.5\columnwidth]{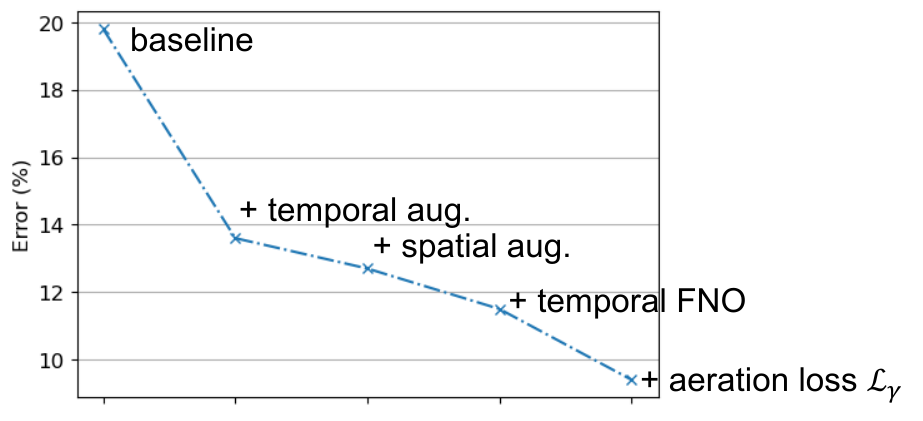}
    \caption{ \small
    {\bf Ablation study results of \method.} The ex vivo percent aeration prediction performance is shown as different components of the proposed model are progressively added. Temporal augmentation and spatial augmentation improve generalization, while the temporal Fourier Neural Operator (FNO) outperforms a ResNet module with equivalent parameters. The addition of aeration loss $\mathcal{L}_\gamma$ further refines predictions, achieving the lowest error.
    }
    \label{fig:supp_ablation}
\end{figure}

In this ablation study, we investigate the impact of different components of our proposed model on the ex vivo percent aeration prediction performance.
\begin{enumerate}
    \item  {\bf Temporal augmentation}. As discussed in Section \ref{sec:architect} of the main paper, temporal masking strengthens the invariance of \method to temporal delay. Empirically, removing temporal augmentation leads to 6.2\% performance loss.
    \item  {\bf Spatial augmentation}. As discussed in Section \ref{sec:architect} of the main paper, spatial masking improves \method's ability to learn lateral correspondence. Empirically, removing spatial augmentation leads to 0.9\% performance loss.
    \item  {\bf FNO module}.  The inclusion of the temporal Fourier Neural Operator (FNO) significantly reduces prediction error compared to replacing it with a ResNet module with equivalent parameters, highlighting the effectiveness of FNO in capturing temporal dynamics. Empirically, replacing FNO with ResNet leads to 1.2\% performance loss.
    \item {\bf The percent aeration loss}. We set the aeration loss's weight $\eta=0$ in Eqn.~\ref{eq:comb_loss}. We show that $\mathcal{L}_\gamma$ further refines the model's predictions by directly optimizing the percent aeration accuracy. Empirically, removing percent aeration loss leads to 2.1\% performance loss.
\end{enumerate}
These components collectively contribute to the performance improvement as shown in ~\cref{fig:supp_ablation}.

\end{appendices}

\end{document}